\documentclass[journal,11pt,draftclsnofoot,onecolumn]{IEEEtran}

\usepackage{amsmath}
\usepackage{amssymb}
\usepackage{amsfonts}
\usepackage{graphicx}

\usepackage{cite}
\usepackage{subfigure}
\usepackage{url}
\usepackage{array}
\usepackage{verbatim}
\usepackage{comment}
\usepackage{stfloats}

\usepackage{color}
\definecolor{red}{rgb}{1,0,0}
\definecolor{blue}{rgb}{0,0,1}
\newcommand{\red}[1]{{\color{red}(#1)}}
\long\def\comment#1{}

\newfont{\bbb}{msbm10 scaled 700}

\newfont{\bb}{msbm10 scaled 1100}
\newcommand{\CC}{\mbox{\bb C}}

\newcommand{\RR}{\mbox{\bb R}}

\newcommand{\EE}{\mbox{\bb E}}

\newcommand{\cv}{{\bf c}}

\newcommand{\ev}{{\bf e}}

\newcommand{\hv}{{\bf h}}

\newcommand{\mv}{{\bf m}}

\newcommand{\sv}{{\bf s}}

\newcommand{\uv}{{\bf u}}

\newcommand{\xv}{{\bf x}}
\newcommand{\yv}{{\bf y}}
\newcommand{\zv}{{\bf z}}
\newcommand{\zerov}{{\bf 0}}
\newcommand{\onev}{{\bf 1}}

\newcommand{\Am}{{\bf A}}
\newcommand{\Bm}{{\bf B}}

\newcommand{\Dm}{{\bf D}}

\newcommand{\Gm}{{\bf G}}
\newcommand{\Hm}{{\bf H}}
\newcommand{\Id}{{\bf I}}
\newcommand{\Jm}{{\bf J}}
\newcommand{\Km}{{\bf K}}
\newcommand{\Lm}{{\bf L}}

\newcommand{\Rm}{{\bf R}}
\newcommand{\Sm}{{\bf S}}
\newcommand{\Tm}{{\bf T}}
\newcommand{\Um}{{\bf U}}

\newcommand{\Xm}{{\bf X}}
\newcommand{\Ym}{{\bf Y}}
\newcommand{\Zm}{{\bf Z}}

\newcommand{\Ac}{{\cal A}}

\newcommand{\Cc}{{\cal C}}

\newcommand{\Ec}{{\cal E}}
\newcommand{\Fc}{{\cal F}}
\newcommand{\Gc}{{\cal G}}

\newcommand{\Nc}{{\cal N}}

\newcommand{\Tc}{{\cal T}}

\newcommand{\thetav}{\hbox{\boldmath$\theta$}}

\newcommand{\xiv}{\hbox{\boldmath$\xi$}}

\newcommand{\Gammam}{\hbox{\boldmath$\Gamma$}}
\newcommand{\Lambdam}{\hbox{\boldmath$\Lambda$}}
\newcommand{\Deltam}{\hbox{\boldmath$\Delta$}}
\newcommand{\Sigmam}{\hbox{\boldmath$\Sigma$}}
\newcommand{\Phim}{\hbox{\boldmath$\Phi$}}

\newcommand{\Thetam}{\hbox{\boldmath$\Theta$}}

\newcommand{\Xim}{\hbox{\boldmath$\Xi$}}

\newcommand{\diag}{{\hbox{diag}}}

\newcommand{\SNR}{{\sf SNR}}

\renewcommand{\Re}{{\rm Re}}
\renewcommand{\Im}{{\rm Im}}
\newcommand{\eqdef}{\stackrel{\Delta}{=}}

\newcommand{\herm}{{\sf H}}

\newcommand{\transp}{{\sf T}}

\begin{document}
\title{Scalable Synchronization and Reciprocity Calibration for Distributed Multiuser MIMO}
\author{\IEEEauthorblockN{R. Rogalin$^1$, O. Y. Bursalioglu$^2$, H. Papadopoulos$^2$, G. Caire$^1$, A. Molisch$^1$,} \\
\IEEEauthorblockN{A. Michaloliakos$^1$, V. Balan$^1$,  and K. Psounis$^1$}
\thanks{$^1$ University of Southern California, Ming-Hsieh Department of Electrical Engineering, Los Angeles, CA 90089, 
email: {\tt rogalin, caire, molisch, michalol, hbalan, kpsounis @usc.edu}}
\thanks{$^2$ DOCOMO Innovations, Inc., 3240 Hillview Avenue, Palo Alto, California 94304,
email: {\tt obursalioglu, hpapadopoulos @docomoinnovations.com}}
}
\maketitle

\vspace{-1cm}

\begin{abstract}
Large-scale distributed Multiuser MIMO (MU-MIMO) is a promising wireless network architecture 
that combines the advantages of  ``massive MIMO'' and ``small cells.''  
It consists of several Access Points (APs) connected to a central server via a 
wired backhaul network and acting as a large distributed antenna system.
We focus on the downlink, which is both more demanding in terms of traffic and more challenging in terms of implementation than the uplink.
In order to enable multiuser joint precoding of the downlink signals, channel state information at the transmitter side is required. 
We consider Time Division Duplex (TDD), where the {\em downlink} channels can be learned from the user {\em uplink} pilot signals, 
thanks to channel reciprocity. Furthermore, coherent multiuser joint precoding is possible only if the APs 
maintain a sufficiently  accurate relative timing and phase synchronization.

AP synchronization and TDD reciprocity calibration are two key problems to be solved
in order to enable distributed MU-MIMO downlink. In this paper,  we propose novel over-the-air synchronization and calibration protocols 
that scale well with the network size.  The proposed schemes can be applied to networks formed by a large number of APs, each of 
which is driven by an inexpensive 802.11-grade clock  and has a standard RF front-end, not  explicitly designed to be reciprocal. 
Our protocols can incorporate, as a building block, any suitable timing and frequency estimator. 
Here we revisit the problem of joint ML timing and frequency estimation and use the corresponding Cramer-Rao bound to evaluate the performance 
of the synchronization protocol. Overall, the proposed synchronization and calibration schemes are shown to achieve sufficient 
accuracy for satisfactory distributed MU-MIMO performance.
\end{abstract}

\begin{keywords}
Distributed multiuser MIMO downlink, cooperative small cells, synchronization, TDD calibration.
\end{keywords}

\section{Introduction}
\label{section:introduction}

The explosive growth of wireless data traffic, spurred by powerful and data-hungry user 
devices such as smartphones and tablets, is putting existing wireless networks under stress 
and is calling for a technology paradigm shift.  
Two recent research trends in wireless networks have attracted significant attention 
as potential solutions:  massive MIMO and very dense spatial reuse.  
The former refers to serving multiple users on the same
time-frequency channel resource by eliminating multiuser interference via 
spatial precoding, using a very large number of  antennas at each base station site \cite{marzetta-massive,Huh11}. 
The latter pertains to the use of small cells \cite{andrews-femtocell-survey}, 
such that multiple short-range low-power links can co-exist on the same time-frequency channel resource 
thanks to sufficient spatial separation.  Distributed Multiuser MIMO (MU-MIMO) unifies these two approaches, 
simultaneously obtaining both multiuser interference suppression through
spatial precoding, and dense coverage by reducing the average distance between transmitters and receivers. 
This is achieved by coordinating a large number of Access Points (APs), 
distributed over a certain coverage region, through a wired backhaul network connected to a 
Central Server (CS), in order to form a distributed antenna system.

While distributed MU-MIMO may be applied in a variety of different scenarios, 
in this work we focus primarily on cost-effective consumer grade equipment and focus particularly on the downlink (DL), 
which is both more demanding in terms of traffic and more challenging in terms of implementation than the uplink (UL). 
We assume inexpensive APs connected to the CS via a conventional wired digital backbone (e.g., Ethernet), 
capable of transporting digital packetized data at sufficient speed from/to the CS, but not able to distribute a common clock (i.e., timing 
and phase information) to the APs, beyond the limitations of standard network synchronization protocols such as IEEE 1588 \cite{ieee-1588}.
In order to achieve large spectral efficiencies through MU-MIMO downlink spatial precoding, 
channel state information at the transmitter (CSIT) is needed. Following the massive MIMO approach  \cite{marzetta-massive,Huh11},  
CSIT can be obtained  from the users' uplink pilots by operating the system in Time-Division Duplexing (TDD) and exploiting
the reciprocity of the radio channels. However, while the propagation channel  from antenna to antenna is reciprocal,\footnote{This statement holds
if the UL and DL transmissions take place within the coherence time and coherence bandwidth of the underlying
doubly-selective channel. Using OFDM and TDD, the propagation channel is effectively reciprocal on each OFDM subcarrier over many 
consecutive OFDM symbols, since the coherence time for typical wireless local area networks
with nomadic users may span $\sim 0.1$s to 1s or more. In fact, $800$ msec is commonly used as the simulation assumption for 801.11ac IEEE standard for WLAN \cite{11ac-800msec}. In WARP software-radio based experiments, we observed the coherence time to be about 1 second.
In contrast, Frequency Division Duplexing schemes have the UL and the DL separated by tens of MHz, well beyond the coherence bandwidth. Therefore, reciprocity for such systems typically does not hold.}  
the transmitter and receiver hardware are not, and introduce an unknown amplitude scaling and phase shift between the 
UL and DL channels \cite{knopp-reciprocity}. 
Furthermore, in order to have all the remote APs behave as a single distributed multiple 
antenna system, the APs must maintain relative timing and phase synchronism throughout the duration of a DL 
precoded slot (see Section \ref{section:model}). 

Since each AP is driven by its own individual clock, 
and the wired backhaul network is not capable of providing
a sufficiently accurate common timing and frequency reference, such synchronization must occur via over-the-air signaling.  
Also, since typical commercial grade radios are not equipped with a built-in self-calibration 
capability,  the non-reciprocal effects of the receiver and transmitter hardware must be compensated explicitly 
via a TDD reciprocity calibration protocol. Note that, although in our treatment we focus on 802.11 grade hardware, the methods we propose are 
platform agnostic. Indeed, the same principles can be applied to any future cost-effective flexible network deployments, where 
practical hardware mismatches need to be eliminated in order to enable coherent reciprocity-based distributed MU-MIMO transmission. 

\begin{figure}[h]
\centerline{\includegraphics[width=12cm]{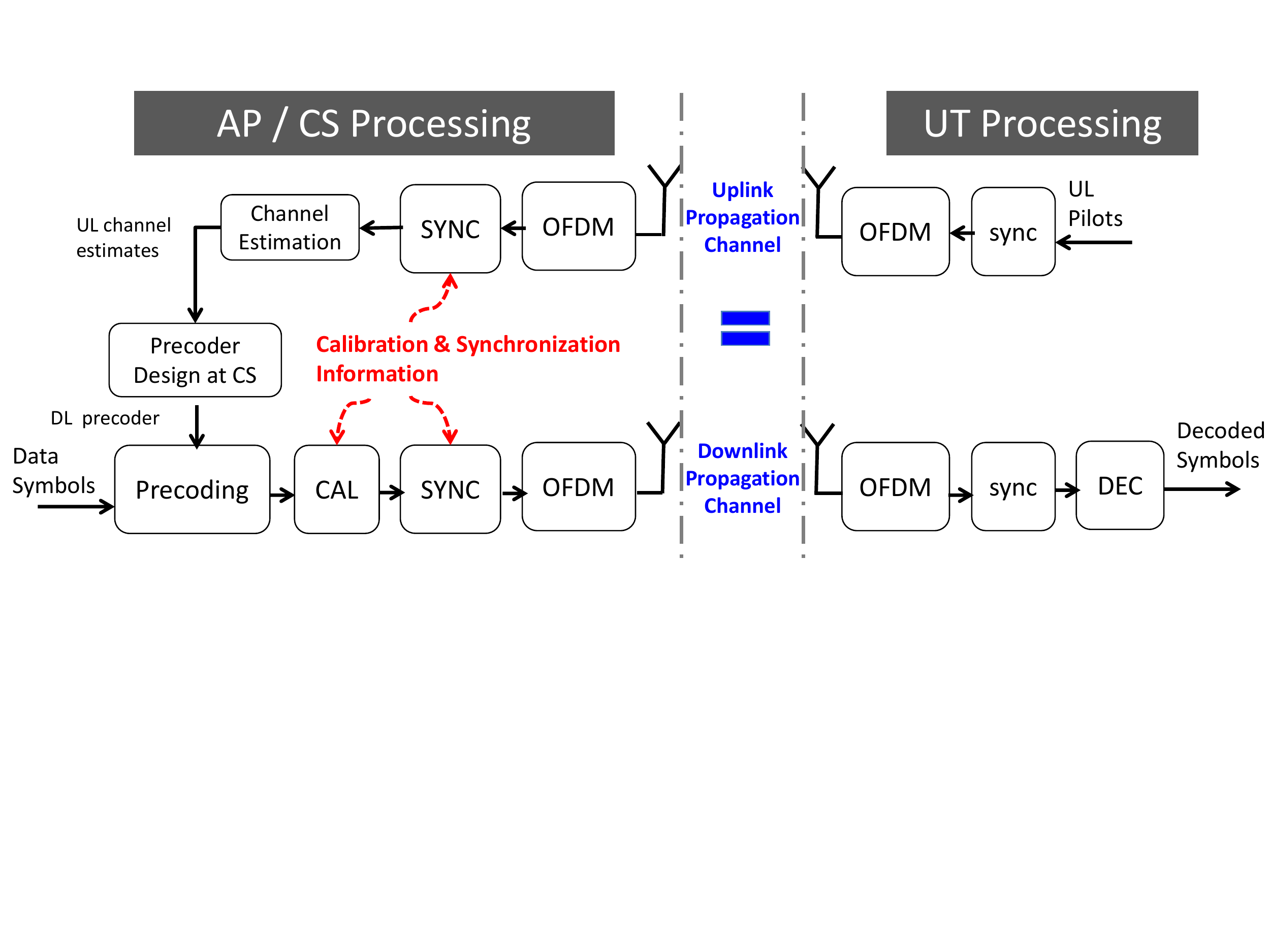}}
\caption{Uplink training and downlink transmission for distributed channel-reciprocity based MU-MIMO.}
\label{fig:UL-DL_fig}
\end{figure}

Fig.~\ref{fig:UL-DL_fig} provides a high-level block diagram that illustrates the operation of channel-reciprocity based MU-MIMO on the hardware, including UL channel training, DL MU-MIMO transmission, and the compensation mechanisms (synchronization and calibration) needed to enable such coherent DL MU-MIMO transmission.  OFDM is assumed throughout this paper, since it is widely used in current WLAN (802.11a/g/n/ac) and cellular (LTE/802.16m) standards \cite{802.11-2012}. 
An UL-pilot/DL-precoded transmission cycle works as follows. At first, UL pilots are transmitted by the User Terminals (UTs).  
After receiving the UL pilots,  each AP individually processes its observations through an OFDM front-end coupled with a standard synchronization block, as used 
in any WLAN current AP implementation. 
Estimates of the UL channels (between each AP and the pilot-transmitting UTs)  are then formed at each AP. These estimates are then sent to the CS through 
the digital backhaul network
and are jointly used at the CS to calculate the MU-MIMO precoding matrix \cite{marzetta-massive,Huh11}. 
In the DL transmission phase, the packets of OFDM encoded symbols destined to the UTs are first precoded using the DL precoding 
matrix. Then, the precoded signal at each AP is calibrated through the ``$\sf {CAL}$'' block, which compensates for the non-reciprocal 
amplitude scaling and phase rotations introduced by the transmitter and receiver AP hardware.\footnote{There are several ways
to apply calibration, some of which are not captured by the diagram in Fig. \ref{fig:UL-DL_fig}.  For example, calibration can be applied to the UL channel 
estimates before the precoding matrix calculation \cite{calibration-ita2013}.  While specific implementations may depend on the specific hardware at hand, all such 
solutions are conceptually equivalent to the scheme in Fig. \ref{fig:UL-DL_fig}.} 
Finally, all the APs forming the distributed MU-MIMO network send simultaneously, on the same time-frequency slot, 
the precoded and calibrated data packets. 

We discuss now the synchronization blocks indicated by ``$\sf {sync}$'' and ``$\sf {SYNC}$'' in Fig. \ref{fig:UL-DL_fig} at the UT and AP side, respectively. 
Synchronization at the UT side (both transmitter and receiver) and at the AP receiver 
takes care of frame and carrier frequency synchronization in order to transmit and receive on the assigned time slots and demodulate 
the OFDM signals with negligible inter-carrier interference (ICI). This can be implemented by completely standard
schemes used in current WLAN technology and forming a well-known and thoroughly studied 
subject \cite{schmidl-cox,vandebeek-ofdm,tufvesson-ofdm,moose-ofdm,kuo-ml-ofdm}. Hence, such blocks need not be treated explicitly in 
this paper.  In contrast, synchronization  at the AP {\em transmitter} side plays a 
critical role in the distributed MU-MIMO architecture, since it needs to compensate for timing misalignment and the relative phase rotation of
the DL data blocks, transmitted {\em simultaneously} by the  jointly precoded APs. Such compensation is non-standard in current network 
implementations and it is {\em not} widely treated in the literature (see the existing-results overview provided below). 
In fact, in most theoretical works on distributed MU-MIMO (see for example 
\cite{Huh-Tulino-Caire,Huh11,ramprashad2009cellular,boccardi2008network,marsch2008base})
perfect transmitter synchronization is {\em assumed} without specifying how this is achieved.  
As shown in Section \ref{sec:sync_impact}, uncompensated synchronization errors at the AP transmitter side cannot be undone 
at the UT receivers, since the jointly precoded signals are mixed up and cannot be individually resampled (for timing) and phase-rotated 
(for carrier frequency offset). For a correct operation of the distributed MU-MIMO architecture it is therefore essential that: 
1) the channel amplitude and phase rotations incurred by UL channel estimation can be compensated at the AP side (TDD calibration), and 2)
the jointly transmitting APs send their precoded data slots with sufficiently accurate timing and carrier frequency coherence. 
This paper focuses on the design of signaling protocols and associated signal processing techniques for the 
$\sf{CAL}$ and $\sf{SYNC}$ blocks at the APs side of Fig. \ref{fig:UL-DL_fig} that accomplish the 
above goals with sufficient accuracy so as to materialize a large fraction of the theoretical capacity gain promised by 
distributed MU-MIMO.

\subsection{Literature Overview}

Both synchronization of multiple transmitters and TDD reciprocity calibration have been treated to some extent in the 
existing literature.  
The synchronization of multiple transmitters for multiple-input single-output (MISO) cooperative beamforming 
is considered in \cite{pottie-txbf,madhow_poor_comm_mag}, and software-defined radio (SDR) implementations of such ideas 
are presented in \cite{mudumbai_ipsn, quitin-wowmom, quitin-globecom}. 
SDR implementations of MU-MIMO have been recently presented by a few research teams (including our own)
in \cite{mobicom2012,airsync-ton,jmb}, where the feasibility of 
over-the-air synchronization with enough accuracy to guarantee near-ideal multiuser interference suppression is demonstrated. 
These works are based on a master-slave protocol, requiring that all APs receive from a single master station broadcasting a beacon signal 
with sufficiently large power and thus limiting the size and topology of the network. Extensions to networks of arbitrary size and topology based on distributed consensus 
protocols are considered in \cite{madhow_consensus, niranjan}.  
The slow rate of convergence, however, makes such schemes impractical for large networks. 
\footnote{For example, \cite{madhow_consensus} reports a convergence 
time of $\sim100$ seconds, widely insufficient to cope with the typical oscillator dynamics of the order of seconds. 
In addition, consensus schemes are targeted to infrastructure-less ad-hoc networks and involve over-the-air message passing between the nodes.
In contrast, for the distributed MU-MIMO application considered in this work, we can effectively and efficiently exploit the presence of the wired backhaul network, 
which is needed anyway to share user data and CSIT across the APs.}
TDD reciprocity was considered in \cite{knopp-reciprocity}, presenting a calibration technique based on 
exchanging pilot signals between transmitter and receiver. In practice, it is much more desirable to design calibration protocols that 
do not assume the presence or collaboration of the UTs, which may be legacy devices not necessarily implementing 
the protocol.  A novel TDD reciprocity calibration method, referred to as ``Argos'', 
was presented in \cite{argos} as part of an SDR implementation of a TDD reciprocity-based massive MIMO base station. 
This method enables calibration by exploiting two-way signaling that involves only base station antennas, exchanging calibration pilot signals 
with a (possibly additional) {\em reference} antenna. It turns out that Argos is very sensitive to the placement of the reference antenna relative to the other 
antennas \cite{argos}.  As a result, this scheme is not readily scalable and is not sufficiently accurate 
to enable large-scale MIMO in a distributed system. 

\subsection{Contributions of this Work}

In this paper we present novel AP synchronization and TDD reciprocity calibration protocols that overcome the drawbacks of 
existing proposals. We assume that the APs form a connected network.~\footnote{Two APs are connected if they can 
receive each other's pilot signals with sufficiently large SNR, where ``sufficiently large'' means above some design threshold determined 
by the system designer.} 
For synchronization, a hierarchical scheme is defined where a subset of ``anchor''  APs is selected at the time 
of network deployment  in order to form a connected cover of the network graph (see Section \ref{section:architecture}). 
The anchor APs exchange pilot signals during special synchronization slots, periodically inserted in the frame structure.
A pilot reuse scheme is used in order to avoid inter-pilot interference.  
The noisy timing and frequency estimates extracted from the pilots are sent to the CS via the wired backhaul, 
and centrally processed by solving a global {\em weighted} Least-Squares (LS) minimization.  
Then, the estimated timing and frequency correction factors are sent back to the anchor APs. 
All other APs get synchronized by their neighboring anchor APs, in a sort 
of ``local master-slave'' scheme. Unlike the master-slave topology, which requires having one AP (the master) with sufficiently 
high SNR to {\em all} other APs (clearly impossible in large networks),  the proposed scheme  has significantly more relaxed 
requirements. First,  it requires a network of  geographically dispersed anchor nodes, whereby each anchor AP is ``connected'' (has sufficiently high SNR) 
to at least another anchor AP, such that the anchor nodes form a connected graph. Second, it requires that the anchor density in the network 
is sufficiently high such that every other AP has at least one anchor in its vicinity, with respect to which it can be synchronized 
(e.g., via a local master-slave protocol).

For calibration we define a spanning connected subgraph of the network and 
have all nodes exchange calibration pilots with their neighbors on this subgraph. 
Noisy complex amplitude estimates of the received pilot signals are
sent to the CS, which finds the calibration coefficients by solving a {\em constrained} LS problem. 
Then, the estimated calibration coefficients are sent back to the APs.  
The complexity of both the weighted LS and the constrained LS  for synchronization and calibration 
is polynomial in the network size and it is easily affordable for networks of practical size. 
In order to extract timing and frequency information from the pilot signals, the APs may use any 
suitable estimator (e.g., \cite{schmidl-cox,vandebeek-ofdm,tufvesson-ofdm,moose-ofdm,kuo-ml-ofdm}).
In this paper,  we consider  joint ML timing and frequency estimation through an unknown multi-path channel and use the corresponding 
Cramer-Rao Bound (CRB) as a good approximation of the estimation MSE achievable 
in practice \cite{kay-single-freq,quinn-threshold}. 

The rest of this paper is organized as follows.  Section \ref{section:model}  presents a simplified but sufficiently accurate 
model for OFDM in the presence of timing, sampling frequency and carrier frequency offsets, and motivates the
need for accurate compensation of these effects in a distributed MU-MIMO scenario. 
Section \ref{section:architecture} discusses the high-level protocol architecture for 
scalable synchronization and TDD reciprocity calibration.  
In Sections \ref{section:synchronization} and \ref{section:calibration}  
we present the details of the synchronization and calibration protocols, respectively, and show their 
effectiveness through system simulations. Conclusions and discussion of future work are pointed out 
in Section \ref{section:conclusion}, and Appendix \ref{crb-ml}
contains the derivation of the joint ML timing and frequency estimation and of the 
associated CRB.

\section{A simple model for OFDM with synchronization errors}
\label{section:model}

Let $f_0$ and $f_s$ be the nominal carrier and sampling frequencies of all nodes of a wireless network, 
respectively. The actual carrier and sampling frequencies of node $i$, indicated by 
$f_{0,i}$ and $f_{s,i}$, respectively, differ from their nominal  values by some {\em deterministic unknown} 
carrier frequency offset (CFO) and sampling frequency offset (SFO), respectively.  
Also, each node $i$ operates according to its local time axis with timing offset (TO) $\tau_i$ 
with respect to a common nominal time axis. 

Motivated by the 802.11 standard \cite{802.11-2012}, we assume that the sampling clock and the RF clock on 
each node hardware are derived from the same (local) oscillator, such that $f_{0,i}/f_{s,i} = \kappa \gg 1$ where 
$\kappa$ is a constant factor that depends only on the hardware design but is independent of $i$. 
Then, we let $f_{s,i} = f_s + \epsilon_i$, where $\epsilon_i \ll f_s$ is the SFO at node $i$. Correspondingly, 
the CFO is given by $\kappa \epsilon_i$. Consistently with the typical accuracy of 802.11 commercial grade hardware, 
we assume that the CFO is much smaller than the signal bandwidth.

Consider the transmission of a sequence of OFDM symbols from node $i$ to node $j$. For the sake of clarity of exposition, 
we assume that each node has a single antenna, although the synchronization and calibration schemes proposed in this work
can be immediately applied to multi-antenna APs.  
We let $\Xm_i[n] = (X_i[n,0], \ldots, X_i[n,N-1])$ denote the $n$-th block of frequency-domain symbols (i.e., the $n$-th OFDM symbol), 
where $N$ is the number of OFDM subcarriers and 
$X_i[n,\nu]$ is the complex baseband  symbol sent on subcarrier $\nu \in \{0,\ldots, N-1\}$ at OFDM symbol time $n$.
The OFDM modulator performs an IFFT, producing the block of $N$ time-domain ``chips''
$\xv_i[n] = {\rm IFFT} (\Xm_i[n]) = (x_i[n,0], \ldots, x_i[n,N-1])$, where $x_i[n,k]$ denotes the $k$-th 
chip of the $n$-th time-domain  block. Each block $\xv_i[n]$ is cyclic-prefixed, forming
$\widetilde{\xv}_i[n] = (\widetilde{x}_i[n,-L], \ldots, \widetilde{x}_i[n,N-1])$, where $L$ is the the length of the cyclic prefix (CP).  
The sequence of cyclic-prefixed blocks is sent to the Digital-to-Analog
Converter (DAC), forming the continuous-time complex baseband signal
\begin{equation} \label{complex-baseband-x}
\widetilde{x}_i (t) = \sum_n  \sum_{k=-L}^{N-1} \widetilde{x}_i[n,k] \Pi((t-\tau_i)/T_{s,i} - n(N+L) - k), 
\end{equation}
where $\Pi(t)$ is the elementary interpolator DAC waveform, assumed for simplicity to be the rectangular pulse
\begin{equation} \label{pipulse}
\Pi(t) = \left \{ \begin{array}{ll}
1 & t \in [-1/2, 1/2] \\
0 & \mbox{elsewhere} \end{array} \right . 
\end{equation}
and where $T_{s,i} = 1/f_{s,i}$ is the sampling interval of transmitter $i$.  The complex baseband signal (\ref{complex-baseband-x}) 
is carrier-modulated in order to produce the transmitted signal $s_i(t) = \Re \left \{ \widetilde{x}_i(t) \exp(j2\pi f_{0,i} t) \right \}$.

Now consider the corresponding receiver operations at node $j$.  Let $h_{i \rightarrow j}(t)$ denote the complex 
baseband equivalent impulse response of the channel between node $i$ and node $j$, including the 
receiver hardware (LNA, Analog-to-Digital Converter (ADC), anti-aliasing filter, etc.).  
Since the CFO between node $i$ and node $j$ is small with respect to the signal bandwidth,  
the complex baseband signal after demodulation can be written as (see also \cite{garcia-oberli})
\begin{eqnarray}
\widetilde{y}_{i \rightarrow j}(t) & = & \left( \sum_n \sum_{k=-L}^{N-1}
\widetilde{x}_i[n,k]  g_{i \rightarrow j} \left ( t - (\tau_i - \tau_j) - (n(N+L) +
k)T_{s,i} \right ) \right) \nonumber \\
& & \times  \exp(j 2\pi (f_{0,i} - f_{0,j}) t),
\label{rx-ij-baseband-ct}
\end{eqnarray}
where we define~\footnote{$(a \otimes b)(t)$ denotes the continuous-time convolution of signals $a(t)$ and $b(t)$.}
 $g_{i \rightarrow j}(t) = (\Pi_i \otimes h_{i \rightarrow j})(t)$, with $\Pi_i(t) =  \Pi(t/T_{s,i})$. 

The ADC at receiver $j$ produces samples at rate $p/T_{s,j}$, where $p \geq 1$ is a suitable oversampling integer factor. 
The resulting discrete-time complex baseband signal,  is given by $\widetilde{y}_{i \rightarrow j}[\ell]  = \widetilde{y}_{i\rightarrow j}(\ell T_{s,j}/p)$. 
Since $T_{s,i}$ and $T_{s,j}$ are both assumed to be very close to the nominal chip interval $T_s$, 
over a single chip, the scale of the time axis at the two nodes 
is essentially identical. However,  by accumulating this small time difference over many OFDM symbols, 
the SFO manifests itself as a timing misalignment of the time axes that grows linearly with 
the OFDM symbol index.  Letting $\ell = m(N+L)p + q$ with $q \in \{-Lp, \ldots, Np-1\}$,  extracting blocks of $p(N + L)$ output samples, 
and assuming  $|\tau_i - \tau_j| \ll T_s L$, i.e.,  that the TO is significantly less than the duration of the CP,~\footnote{As it will be clear from Section \ref{section:architecture}, the synchronization protocol that we propose in this paper makes sure that this assumption is satisfied.} we arrive at the following channel model without inter-block interference between consecutive OFDM symbols: for $q \in \{ 0, \ldots, pN-1\}$
\begin{eqnarray}\label{rx-approx}
\widetilde{y}_{i \rightarrow j} [m,q] & =  &  \left( 
\sum_{k =-L}^{N-1} \widetilde{x}_i[m,k]  
g_{i \rightarrow j}  ( (q - kp) T_s/p   - (\tau_i - \tau_j) - m (N+L)(T_{s,i} - T_{s,j})  ) \right)
\nonumber \\
& & \times \exp(j 2\pi (f_{0,i} - f_{0,j}) T_{s}( m(N+L) + q/p)).
\end{eqnarray}
Since $|f_{0,i} - f_{0,j}| T_s \ll 1$, the phase rotation across the subcarriers of a single OFDM symbol is negligible. Hence, we can 
drop the dependence on $q$ from the argument of the exponential in (\ref{rx-approx}). 
This is equivalent to approximating the instantaneous phase of the carrier term in (\ref{rx-approx}) as piecewise constant on each
OFDM symbol of duration $(N+L)T_s$, and taking into account only the phase increment in discrete steps from one OFDM symbol to the next. 
This approximation is the time-domain equivalent of neglecting ICI, which is
indeed a valid approximation when the CFO is much smaller than the OFDM subcarrier separation. 
\footnote{This assumption is validated by the oscillator requirements of the 802.11-2012 standard, which allows 20 ppm frequency error. For a $f_o = 2.45$ GHz system,  this corresponds to a maximum CFO value of $98$ kHz, which is much less than $312.5$ KHz, the subcarrier spacing in the 802.11 standard.}
Notice also that we make this approximation here for the sake of obtaining a simple model for deriving the proposed CFO/SFO/TO compensation
scheme. However,  in our simulations we work with the full signal model (\ref{rx-approx}) such that the effect of residual ICI is taken into account. 

The received discrete-time $m$-th signal block after CP removal can thus be expressed by 
\begin{eqnarray} \label{post-cp}
y_{i \rightarrow j} [m,q] & \approx & \left ( 
\sum_{k =-L}^{N-1} x_i[m, k \; {\rm mod} \; N]   
g_{i \rightarrow j}  ( (q - kp) T_s/p   -  (\tau_i - \tau_j) - m (N+L)(T_{s,i} - T_{s,j})  ) \right)
\nonumber \\
& & \times \exp(j 2\pi (f_{0,i} - f_{0,j}) T_{s}(N+L) m), \;\;\; q = 0, \ldots, pN-1.
\end{eqnarray}
The first term in the above product is easily recognized to be  the cyclic convolution of 
the up-sampled sequence $\{x_i[m,k] : k=0, \ldots, N-1\}$ (with insertion of $p-1$ zeros in between each sample), with the 
discrete-time impulse response
\begin{equation} 
g_{i \rightarrow j}[m,q] = g_{i \rightarrow j}  ( q T_s/p - (\tau_i - \tau_j) - m (N+L)(T_{s,i} - T_{s,j}) ), \;\;  \text{for }q = 0, \ldots, pN-1.
\end{equation}
We wish to obtain a frequency-domain representation of this cyclic convolution. To this purpose, 
notice that $g_{i \rightarrow j}[m,q]$ is obtained by sampling at rate $p/T_s$ the impulse response
$g_{i\rightarrow j}(t)$ delayed by  $(\tau_i - \tau_j) + m (N+L)(T_{s,i} - T_{s,j})$. 
Using the well-known {\em spectral folding} relationship between 
the continuous-time Fourier transform and the discrete-time Fourier transform of the corresponding sampled signal  
\cite{oppenheim}, and after some algebra we find the discrete-time Fourier transform of $g_{i\rightarrow j}[m,q]$ in the form 
\begin{eqnarray}
G_{i \rightarrow j}(m,\xi) & = & \sum_{q=0}^{pN-1} g_{i \rightarrow j}[m,q] e^{-j2\pi \xi q} \nonumber \\
& = & \frac{p}{T_{s}} \sum_{\ell} \Gc_{i \rightarrow j} \left (\frac{\xi -
\ell}{T_{s}/p} \right )  \times 
\exp\left(-j2\pi (\xi - \ell) \frac{(\tau_i - \tau_j) + m (N+L)(T_{s,i} - T_{s,j})}{T_{s}/p}\right), \nonumber \\
& & 
\end{eqnarray}
where $\Gc_{i \rightarrow j}(f) = \int g_{i\rightarrow j}(t) e^{-j2\pi ft} dt$ is the continuous-time Fourier
transform of $g_{i\rightarrow j}(t)$, and where $\xi \in [-1/2,1/2]$ and $f\in \RR$  
denote the frequency variables for discrete-time and continuous-time signals, respectively. 

Usually, systems are designed such that $N \gg L \gg 1$ and the receiver sampling
frequency $p/T_s$ is large enough in order to avoid significant spectral folding. This means that 
only the term for $\ell = 0$ is significant in the (discrete-time) frequency interval $\xi \in [-1/(2p), 1/(2p)]$, 
which is  the interval of interest containing the $N$ OFDM subcarriers of the transmitted signal. 
Hence, after applying the DFT and taking the samples corresponding to the set of discrete frequencies
$\{\nu/(pN) : \nu = -N/2, \ldots, N/2-1\}$, straightforward algebra (omitted for the sake of brevity) yields
the OFDM frequency-domain demodulated signal in the form
\begin{equation} \label{terms}
Y_{i \rightarrow j}[m, \nu] = H_{i \rightarrow j}[\nu] X_i[m, \nu] E_{i,j}[m,\nu],
\end{equation}
where
$H_{i \rightarrow j}[\nu] = \frac{p}{T_{s}} \Gc_{i \rightarrow j} \left
(\frac{\nu}{NT_s} \right )$,
and where 
\begin{eqnarray}
E_{i,j}[m,\nu] & = &  \exp\left ( - j2\pi  \left [\frac{\nu}{N} \right ]
(\mu_i - \mu_j) \right ) \label{Eterm-time} \nonumber \\
& & \times  \exp \left ( j2\pi \left [\frac{\nu}{N} \right ]  ( \delta_i -
\delta_j) m \right )  \label{Eterm-sampling} \nonumber\\
& & \times \exp \left ( j2\pi (\Delta_i - \Delta_j) m \right ),
\label{Eterm-carrier} \nonumber
\end{eqnarray}
with $\nu \in \{-N/2, \ldots, N/2-1\}$ and where  we define the normalized TO, SFO and CFO as
\begin{equation} \label{norm-offset}
\mu_i \eqdef f_s \tau_i, \;\; \delta_i \eqdef (N+L) T_s \epsilon_i, \;\; \Delta_i \eqdef \kappa \delta_i,
\end{equation}
respectively.

In (\ref{terms}) the term $H_{i \rightarrow j}[\nu] X_i[m, \nu]$ represents what we would expect from an ideal system 
without synchronization errors while the combined effects of the TO, CFO and SFO are captured by the multiplicative phase 
rotation term $E_{i,j}[m,\nu]$.\footnote{This model has been validated experimentally by the authors 
using the WARP software radio platform, as documented in \cite{mobicom2012,airsync-ton}, and it is found to be accurate within 
the typical errors of 802.11 legacy APs.}

\subsection{Impact of Synchronization Errors on Distributed MU-MIMO}
\label{sec:sync_impact}

We consider a network formed by UTs
$k = 1,\ldots, N_u$ served in the DL by APs $i = 1, \ldots, N_a$, 
using distributed MU-MIMO. 
From the analysis developed above, it is immediate to show that the received signal $1 \times N_u$ vector
at the UT receivers, at OFDM symbol $m$ and subcarrier $\nu$, is given by 
\begin{equation}\label{bb-model}
\Ym[m,\nu] = \Xm[m,\nu] \Phim[m,\nu]   \Hm[m, \nu] \Thetam[m,\nu]  + \Zm[m,\nu],
\end{equation}
where $\Hm[m, \nu]$ is the $N_a \times N_u$  channel matrix with $(i, k)$-th
element $H_{i,k}[m,\nu]$,  $\Xm[m,\nu]$ is the $1 \times N_a$ vector of frequency domain symbols
transmitted by the $N_a$ APs simultaneously,  and $\Zm[m,\nu]$ is the corresponding $1 \times N_u$ vector 
of Gaussian noise samples, i.i.d. $\sim \Cc\Nc(0,N_0)$. 
In general, the MU-MIMO precoded signal vector is given by 
\begin{equation} \label{precodedX}
\Xm[m,\nu] = \Um[m,\nu] \Gm[m,\nu], 
\end{equation}
where $\Um[m,\nu]$ is the $1 \times N_u$ vector of time-frequency encoded data symbols to be sent to the $N_u$ UTs, 
and $\Gm[m,\nu]$ is the $N_u \times N_a$ MU-MIMO precoding matrix for OFDM symbol $m$ and subcarrier $\nu$. 
In this work we consider linear Zero-Forcing Beamforming (ZFBF) \cite{caire2010multiuser,Huh11,argos}
and {\em conjugate beamforming} \cite{marzetta-massive,Huh11,argos}, the details of which are given later on in this section.

The two matrices  $\Thetam[m,\nu]$ and $\Phim[m,\nu]$ are diagonal of dimension $N_u \times N_u$ and $N_a \times N_a$, 
respectively, with diagonal elements given by\footnote{The equations derived above apply both when the nodes are APs and UTs. To distinguish APs from UTs we use tildes over the UT variables. In this case $\widetilde{\mu}_k, \widetilde{\delta}_k$ and $\widetilde{\Delta}_k$ denote the 
normalized TO, SFO and CFO of UT $k$ with respect to the nominal time and frequency references.}
\begin{equation} 
\theta_{k,k}[m,\nu] = \exp\left (j2\pi  \left [\frac{\nu}{N} \right ] \widetilde{\mu}_{k}
\right ) 
\exp \left (- j2\pi \left [\frac{\nu}{N} \right ] \widetilde{\delta}_{k} m \right )  \exp
\left (- j2\pi \widetilde{\Delta}_{k} m \right ), 
\end{equation}
and
\begin{equation} \label{factor-to-be-corrected}
\phi_{i,i}[m,\nu] = \exp\left (- j2\pi  \left [\frac{\nu}{N} \right ] \mu_{i}
\right ) 
\exp \left (j2\pi \left [\frac{\nu}{N} \right ] \delta_{i} m \right )  \exp
\left ( j2\pi\Delta_{i} m \right ).
\end{equation}
The effect of $\Thetam[m,\nu]$ can be compensated at each UT (see $\sf{sync}$ blocks  at UTs in Fig.~\ref{fig:UL-DL_fig}) by standard 
pilot-aided or data-aided timing and frequency synchronization techniques suited to OFDM (see \cite{schmidl-cox,vandebeek-ofdm,tufvesson-ofdm,moose-ofdm,kuo-ml-ofdm}). On the other hand, the presence of  $\Phim[m,\nu]$ in (\ref{bb-model}) between the precoded transmit vector $\Xm[m,\nu]$ 
and the  channel matrix $\Hm[m,\nu]$ yields a degradation of the MU-MIMO precoding performance that cannot be 
undone by UT processing.  As anticipated in Section \ref{section:introduction}, in this paper we focus on the compensation of the term 
$\Phim$ at the APs transmitter side.   
As shown in Fig.~\ref{fig:UL-DL_fig},  our approach includes the use of a central server ($\sf{CS}$) module and $\sf{SYNC}$ blocks, responsible for carrying out the synchronization at the transmitter side.  The MU-MIMO precoding matrix  $\Gm[m,\nu]$ in (\ref{precodedX}) 
is constant over time throughout the transmission block (i.e., it is independent of $m$), 
and it is calculated from the noisy estimate of the  ``nominal" channel  matrix $\Hm[0,\nu]$ obtained by the UL pilots at the beginning 
of the precoding block  (here indicated as OFDM symbol $m = 0$). Hence, the precoder is ignorant of the matrix $\Phim[m,\nu]$ 
unless  TO, SFO and CFO are explicitly taken into account. In the case of ZFBF, we let
$\Gm[m,\nu] = \Lambdam[0,\mu] (\Hm^\herm[0,\nu] \Hm[0,\nu])^{-1} \Hm^\herm[0,\nu]$ for all $m$, 
and in the case of conjugate beamforming we let  $\Gm[m,\nu] =  \Gammam[0,\nu] \Hm^\herm[0,\nu]$, where $\Lambdam[0,\nu]$ and $\Gammam[0,\mu]$ 
are diagonal matrices that ensure that the rows of $\Gm[m,\nu]$ have unit norm. 
Notice also that the ZFBF precoder requires $N_u \leq N_a$.

\begin{figure}[h]
\centerline{\includegraphics[width=10cm]{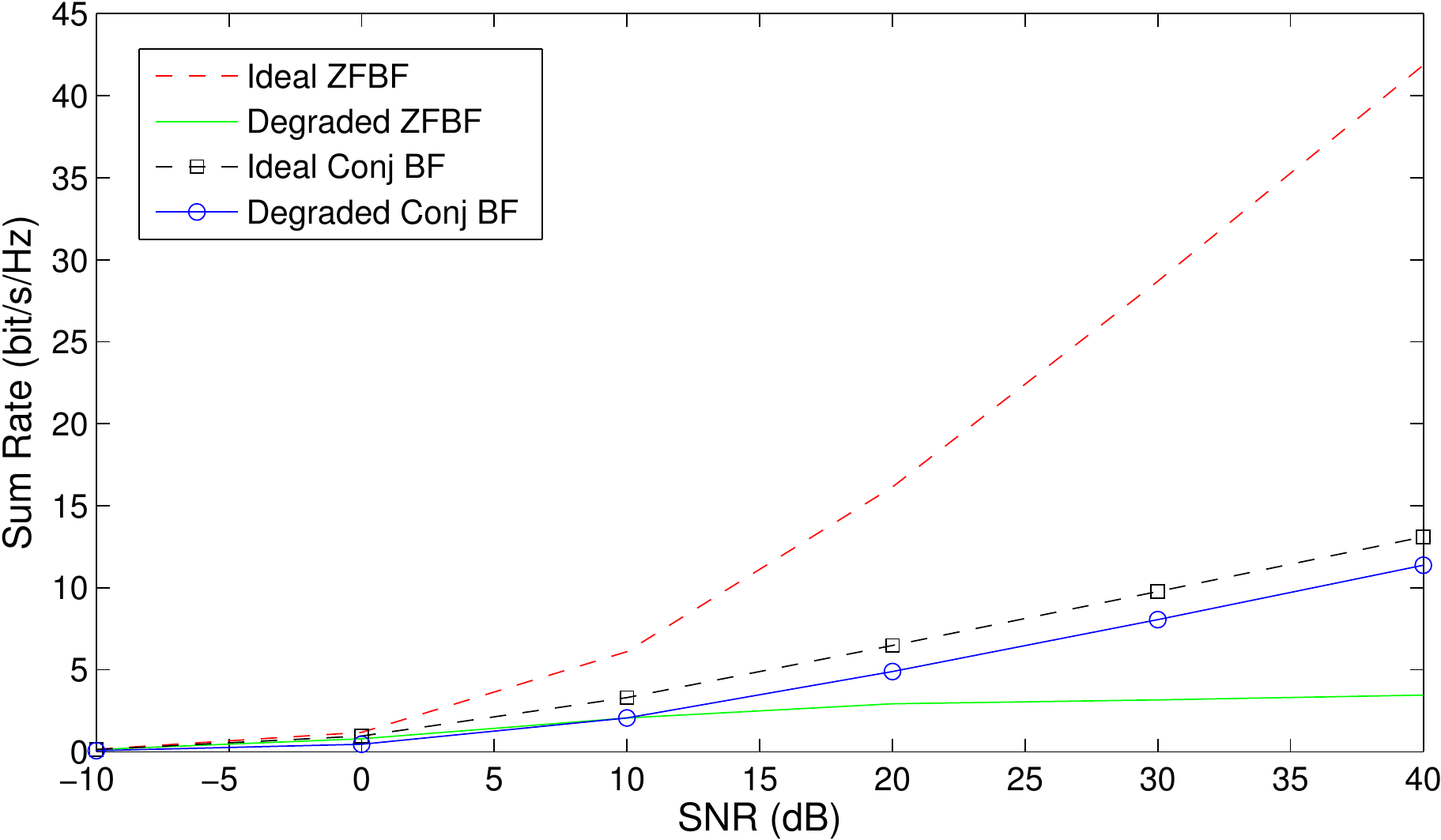}}
\caption{Achievable average rates for a $4\times 4$ distributed MIMO systems in ideal synchronization conditions and 
degraded by synchronization impairments (uncompensated free running oscillators).}
\label{fig:achievablerates}
\end{figure}

In order to motivate the need for the synchronization protocol proposed in this paper, 
we examine the performance degradation due to typical uncompensated SFO and CFO between the APs.  
We use lower bounds on the achievable sum-rates as a performance metric, 
obtained assuming an i.i.d. equal-power Gaussian coding ensemble, 
where the UTs treat the residual interference due to imperfect zero-forcing as noise.\footnote{Quantifying the system performance in terms of 
achievable rates is now common practice in modern applied communication works (e.g., see \cite{argos,mobicom2012,airsync-ton,jmb}) 
and it is much more significant than evaluating traditional performance metrics such as self-interference variance, estimation MSE, or uncoded BER, 
for which we are typically unable to quantify their impact on the ultimate system performance, which makes use of powerful 
modulation and coding schemes.} 
We assume DL blocks of $M = 60$ OFDM symbols, typical of 802.11 \cite{802.11-2012}. 
In order to focus on the impact of synchronization errors only, $\Hm[m,\nu]$ is assumed constant with respect to the time 
index $m = 0,\ldots, M-1$ over each block  and, optimistically, 
we assume ideal TDD reciprocity and noiseless channel estimation. 
Provided that  the relative TO between APs is within the length of the CP, the terms  $\exp\left (- j2\pi  \left [\frac{\nu}{N} \right ] \mu_{i} \right )$ are 
automatically included as part of the channel estimated from the UL pilot symbols.\footnote{It can be readily shown that it  takes about 1 second for two completely uncompensated clocks to fall out of sync by more than the CP,  if they are off by worst case frequency offsets dictated by 802.11. Hence, synchronization performed at intervals 1 second apart or faster suffices.} 
Under these assumptions, the CS computes its MU-MIMO precoding matrix based on the channel matrix $\Phim[0,\nu] \Hm[0,\nu]$ and uses it 
throughout the whole DL block comprising $M$ symbols.  Therefore, because of the time-varying matrix $\Phim[m, \nu]$, the precoder 
is more and more mismatched as $m$ increases. 
Fig.~\ref{fig:achievablerates} shows the achievable rates obtained by Monte Carlo simulation
assuming $\Hm[0,\nu]$ with i.i.d. elements $\sim \Cc\Nc(0,1)$ (normalized independent Rayleigh
fading) of a network with $N_a = N_u = 4$, SFO $\epsilon_i$ i.i.d. across the users 
and the DL blocks, uniformly distributed over $[-\epsilon_{\max}, \epsilon_{\max}]$, 
with  $\epsilon_{\max}=800\text{ Hz}$ (20 ppm frequency error), and $f_s=40\text{ MHz}$ from the 802.11 specifications \cite{802.11-2012}. 
The OFDM modulation has parameters $N=64$ and $L=16$.  The achievable rate shown in Fig.~\ref{fig:achievablerates}  
is the average rate across a frame of 60 OFDM symbols. Since the system loses synchronism progressively across each block, 
the achievable rate rapidly degrades as the OFDM symbol $m$ increases, such that the average performance 
is severely degraded.  
As a comparison, we also show the performance of an ideally synchronized system (zero CFO/SFO) 
and the performance of a MISO cooperative beamforming scheme employing  conjugate beamforming  (denoted by Conj BF in the figure)
to a single user \cite{argos,madhow_poor_comm_mag}, 
with and without frequency offsets. 
Notice that MISO cooperative beamforming  suffers much less from the lack of synchronization, and 
significantly outperforms the mismatched MU-MIMO ZFBF at high SNR.  
Thus the motivation of this work is to provide the APs 
with sufficiently accurate estimation of their SFO/CFO such that the large spectral efficiency promised by 
distributed MU-MIMO is effectively realized in practice. 

\section{System Architecture}
\label{section:architecture}

At the network setup phase, each AP discovers its neighbor APs and measures the associated 
propagation delays. Since the APs are placed in fixed positions, the network structure (neighboring APs) and path delays 
change very slowly in time and it is sufficient to repeat such procedure at a very slow time scale (e.g., of the order of tens of minutes or slower) 
to safely assume that these parameters are known. 

Calibration for TDD reciprocity (Section \ref{section:calibration}) depends on the complex scaling factors  
introduced by the modulation/demodulation hardware (amplifiers, filters, ADC/DACs), that change 
slowly in time.  Experimental SDR implementation shows that calibration must be repeated on a time-scale on the order 
of 10 min \cite{argos} for a system where all of the APs share a common clock.  
In contrast, synchronization must track the variations of the clock frequencies. 
While in this paper we model the frequency errors $\{\epsilon_i\}$ as unknown 
constants, in practice they depend on temperature, and fluctuate around their nominal 0 value  with  a coherence time on the order of 1s. 
Hence synchronization must be performed frequently enough such that the errors do not significantly degrade the system throughput.

\begin{figure}[ht]
\centerline{\includegraphics[width=14cm]{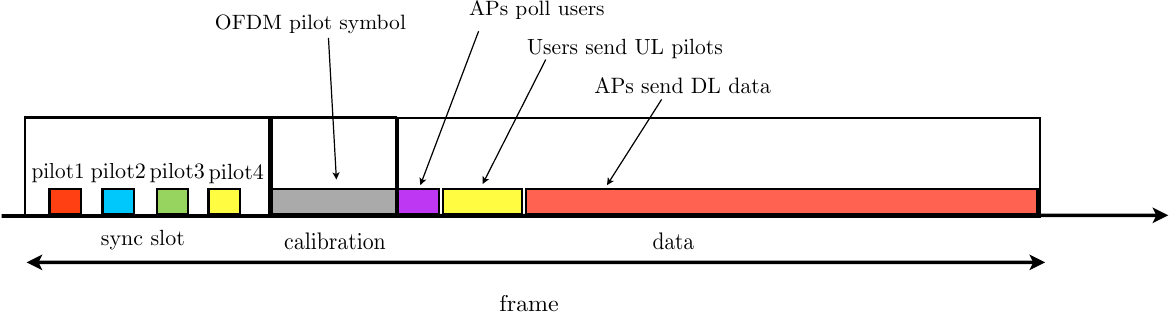}}
\caption{Concept of the frame structure for the proposed distributed MU-MIMO architecture.
The synchronization slot contains several pilot signals, orthogonal in time, and separated by guard intervals. 
The pilot signals are assigned to the anchor APs according to a graph coloring scheme, similarly to what is done with frequency reuse in cellular systems. 
In this example, we have 4 orthogonal pilot signals. The  calibration slot comprises standard OFDM pilot symbols.  
During this slot, each AP transmits pilots over one or more time-frequency slots, allowing it to calibrate itself with other APs in its vicinity. 
Again, pilot patterns orthogonal in time-frequency are allocated to the APs and reused across the network according to a graph coloring scheme.
The relative size of the sync and calibration slots with respect to the data slots in this figure do not reflect the actual duration in a real implementation. 
As a matter of fact, the protocol overhead is very small for typical channel coherence times in a WLAN nomadic users scenario.}
\label{fig:superframe}
\end{figure}

In a distributed MU-MIMO system where APs do not share a common clock,
the very small residual frequency errors due to non-ideal synchronization cause slow relative drift among the different AP phasors. 
As these uncompensated phase rotations accumulate over time,  and since they affect the extent of end-to-end reciprocity in the channel, 
calibration needs to be performed at the same rate as synchronization. 
The proposed protocol makes use of a frame structure comprised of synchronization, 
calibration and DL data slots, as shown in Fig.~\ref{fig:superframe} where synchronization and calibration take place back to back.

Let $\Tc = \{i : i  = 1, \ldots, N_a\}$ denote the set of APs. We define a directed network graph with 
vertices $\Tc$ and edges $\Ec = \{(i,j)\}$ for all ordered pairs of APs $i, j$ such that the {\em average} SNR of the 
transmission $i \rightarrow j$  is larger than some suitably defined threshold. Since the distance-dependent pathloss 
is symmetric with respect to the AP indices, if $(i, j) \in \Ec$ then also $(j, i) \in \Ec$. 
Without loss of generality, we assume that the graph $(\Tc, \Ec)$ is connected.\footnote{Otherwise, the same algorithms can be applied 
independently to each connected component.} 

\begin{figure}[ht]
\centerline{\includegraphics[width=10cm]{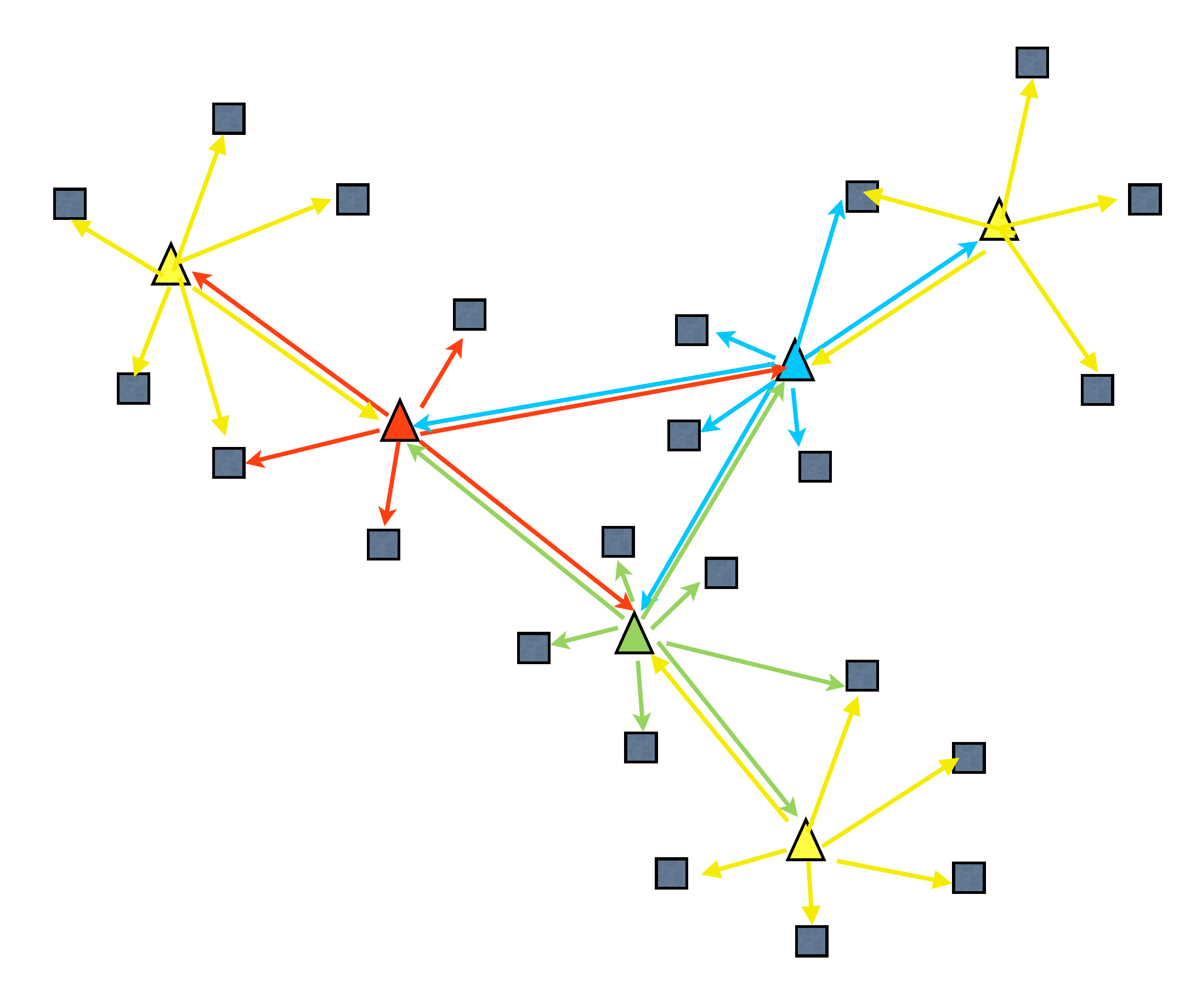}}
\caption{Network graph with 6 anchor nodes (triangles) reusing 4 orthogonal pilot bursts 
(identified by different colors, as in Fig.~\ref{fig:superframe}). The non-anchor APs do not transmit sync pilots, and listen to their neighbor 
anchor node(s) pilot burst  in order to correct their timing and frequency reference.}
\label{fig:anchor-nodes}
\end{figure}

For the synchronization protocol, we define $\Ac \subseteq \Tc$ to form a {\em connected cover} \cite{zhang-gao-wu}, i.e., 
the subgraph formed by the APs $i \in \Ac$ and their associated edges $\Ec(\Ac) = \{ (i,j) : i, j \in \Ac\}$ 
is connected and any other AP $i' \notin \Ac$ has at least one neighbor in $\Ac$. 
The APs in $\Ac$ are referred to as ``anchor'' nodes.  
Fig.~\ref{fig:anchor-nodes} shows a network with 6  anchor nodes.  
In a practical system deployment, we may imagine that  each cluster corresponds to a geographically isolated area (e.g., a building) 
and the anchor nodes are placed in special positions (e.g., roof-top), such that they can exploit stronger propagation 
conditions (e.g., line of sight) to each other.  
During the network deployment phase, the network designer or some self-organizing 
scheme should find a good set of anchor APs. Such optimization, however, is beyond the scope of this paper 
and is left for future investigation. 

The synchronization slots are formed by a number of {\em pilot bursts} (sequences of time-domain chips) separated by 
guard intervals.  We assume that a rough timing synchronization is provided 
through the wired backhaul network using a coarse frame synchronization protocol such as 
IEEE 1588 \cite{ieee-1588}. As a result, the APs  know where to look for the pilot bursts up to an accuracy of 
a few OFDM symbols,  corresponding to the guard intervals inserted between the pilot bursts. 
Pilot bursts are assigned to the anchor nodes. We require that each anchor node is able to receive
the pilot bursts from its neighboring anchor nodes without significant co-pilot interference.
This is possible by finding an $L(1,1)$-labeling (see \cite{calamoneri2006h} and references therein) 
of the subgraph $(\Ac, \Ec(\Ac))$, and associating 
pilot bursts to the labeling colors. An $L(1,1)$-labeling of a graph is a coloring scheme for which all neighbors of any node $i \in \Ac$ 
have distinct colors. Using a number of orthogonal pilot bursts larger or equal to the 
number of colors of the $L(1,1)$-labeling, guarantees that each anchor AP is able to receive  pilots from its neighbors without pilot collisions.
Notice that the orthogonal pilots are re-used across the network. For example, the 6 anchor nodes of Fig.~\ref{fig:anchor-nodes} require only 4 orthogonal pilots.
More generally, when $(\Ac, \Ec(\Ac))$ is a cycle, if $|\Ac|$ is a multiple of 
3, then 3 colors are sufficient, while if $|\Ac|$ is not multiple of 3 then 4 colors are needed. 
If $(\Ac, \Ec(\Ac))$ is a clique (fully connected graph), then $|\Ac|$ is the necessary and sufficient number 
of colors for an $L(1,1)$ labeling,  and if $(\Ac, \Ec(\Ac))$ is a tree, then it is known that $d_{\max} + 1$ colors are necessary 
and sufficient,  where $d_{\max}$ is the maximum node degree. 
The non-anchor APs in $\Tc - \Ac$ do not transmit pilot bursts, and just listen to the anchor nodes. 
Eventually, at the end of a synchronization slot, each anchor 
node has received a distinct and non-overlapping pilot burst from each of its neighbors 
in $(\Ac, \Ec(\Ac))$,  and all non-anchor nodes have received one or more pilots from their neighboring 
anchor nodes.  The estimates collected by the nodes are sent through the wired backhaul network to the CS, 
which computes correction factors as explained in Section \ref{section:synchronization}.

The calibration slot is organized in a similar way, with the only difference being that in this case {\em all} nodes have to send 
and receive a pilot signal, as explained in Section \ref{section:calibration}. 
If the network is formed by sufficiently isolated clusters, calibration can be organized in a hierarchical way such that 
all clusters calibrate in parallel, and then a second round of calibration is run between the anchor nodes (one per cluster) in order to
calibrate the whole network (see \cite{calibration-ita2013}). For example, such hierarchical schemes prove to be very useful in the presence of
multiple antenna APs, where the groups of antennas of each AP, driven by a common clock, can be calibrated ``locally'' and in parallel 
using the scheme of \cite{argos}, and a second level of calibration is used to set the relative calibration coefficients between the APs.
This again makes the proposed scheme to work in even very large networks without the requirement of a ``master" node or calibration antenna 
connected with high SNR to all other nodes or antennas. 

Thanks to the periodically repeated computation of the timing, frequency and calibration correction factors, 
the APs are time/frequency synchronous and calibrated and can operate (up to residual errors) as a large distributed  
and coherent multi-antenna system.  Both synchronization and calibration must remain sufficiently accurate over the duration of 
the data transmission slot in the frame, and until the next synchronization and calibration 
slot (see Fig.~\ref{fig:superframe}).

Notice that for network graphs of bounded degree (the typical case in 
wireless networks with propagation pathloss) the number of orthogonal pilots does not need to grow with the network size. 
In fact, the orthogonal pilot reuse of the proposed protocol is analogous to frequency reuse in cellular networks, 
where a finite (typically small) number of subchannels are reused throughout 
the network irrespectively of the number of cells, which can be arbitrarily large. 
Hence, the proposed synchronization and calibration protocols are 
{\em scalable}, in the sense that the pilot overhead is {\em independent of the network size}.

The data slots can be arranged in the classical way widely proposed in TDD-based MU-MIMO literature 
\cite{marzetta-massive,Huh11,mobicom2012,airsync-ton,argos,caire2010multiuser}:  the APs send a request for UL pilots. 
The polled users respond with their mutually orthogonal UL pilots. 
Then, the APs send the received UL pilot burst to the CS which estimates the UL channel, computes the corresponding DL 
MU-MIMO precoding matrix and sends the precoding coefficients (columns of the precoding matrix) along with the encoded data packets back to the APs for transmission.
Finally, the APs pre-code the user data packets according to (\ref{precodedX}), and compensate for 
the TDD calibration factors (see Section \ref{section:calibration}) and phase rotation factors due to timing and frequency offsets
(see Section \ref{section:synchronization}) before joint DL transmission.

\section{Synchronization} \label{section:synchronization}

In order to overcome the impairments discussed in Section \ref{section:model},
we observe that the effects of the error term (\ref{factor-to-be-corrected}) can be undone by each AP $i$ transmitter 
by multiplying the transmitted frequency domain symbols $X_{i}[m,\nu]$ by the time and frequency dependent phase rotation factor
\begin{equation} \label{symbol-rotation}
P_{i}[m,\nu] =  \exp \left ( - j2\pi \left ( 1  + \frac{1}{\kappa} \left [\frac{\nu}{N} \right ] \right )  \Delta_{i}^{corr} m \right ),
\end{equation}
and by adjusting the timing reference (origin of the transmitter time axis) by $\mu_{i}^{corr}$. 
The timing and frequency corrections $\Delta_{i}^{corr}, \mu_{i}^{corr}$ are provided by the CS through the 
low-complexity centralized computation illustrated in the rest of this section. 

We should also notice that a symbol phase rotation by $P^*_{i}[m,\nu]$, i.e., the complex conjugate of the term in (\ref{symbol-rotation}),
must be applied to signals {\em received} at the APs during the UL training from the users, in the data slots. 
In this way, all APs use  the same time and frequency reference for the 
calculation of the DL channel matrix for  MU-MIMO precoding.  Performing such corrections during UL training and DL transmission is consistent with the diagram in 
Fig. \ref{fig:UL-DL_fig}, showing  $\sf{SYNC}$ block modules in both UL and DL AP/CS processing.

In addition to the baseband symbol rotation as modeled in Section \ref{section:model}, the SFO produces a contraction or dilation of the  
time axis of each AP. While this is a negligible effect over a few tens of OFDM symbols (typical duration of a data slot), it accumulates over the slots such that at some point the OFDM symbol misalignment  between between different APs becomes larger than the OFDM CP, thus producing inter-block interference and not only relative phase rotations.  However, for typical values of SFO consistent with the 802.11 standard specifications, 
we have checked that the duration for which this symbol misalignment becomes significant with respect to the OFDM CP length
is much larger than the frame length at which the synchronization protocol must be repeated. Therefore, thanks to the TO correction, 
the timing axis relative shift is reset at each frame and the SFO can be effectively compensated by the proposed baseband symbol rotation scheme.


Each anchor node collects pilot bursts from all its neighboring anchor nodes and 
produces estimates of the TO and CFO with respect to its neighbors. While
any suitable estimation scheme can be used for this purpose, in Appendix \ref{crb-ml} we derive the 
joint ML timing and frequency estimator and of the corresponding  CRB in the case, relevant to our scenario, 
where the pilot signal is observed through a multipath channel with known path delays and unknown path  coefficients. 

In this section we focus on the centralized estimation of the correction factors 
$\Delta^{corr}_i$ and $\mu^{corr}_i$ from the noisy CFO and TO estimates
\begin{equation} \label{f-estimates}
\widehat{\delta \Delta}_{j \rightarrow i} = \Delta_{j} - \Delta_{i} + z_{j \rightarrow i}, 
\end{equation}
and
\begin{equation} \label{tau-estimates}
\widehat{\delta \mu}_{j \rightarrow i} = \mu_{j} - \mu_{i} + w_{j \rightarrow i},
\end{equation}
for all $i \in \Ac$ and for all $j \in \Ac(i)$,\footnote{In the following, $\Ac(i)$ denotes the neighborhood of AP $i$ in the  subgraph $(\Ac, \Ec(\Ac))$.}
where $z_{j \rightarrow i}$  and $w_{j \rightarrow i}$ are 
estimation errors with variance equal to the corresponding estimation MSE. 
As already noted, if $j \in \Ac(i)$ then $i \in \Ac(j)$. Therefore, these measurements always come in pairs.  

The CS collects all these estimates and computes the correction terms $\Delta_{i}^{corr}, \mu_{i}^{\rm corr}$. 
In the absence of estimation errors we would have $\widehat{\delta \Delta}_{j \rightarrow i} = \Delta_{j} - \Delta_{i}$. Hence, a reasonable approach consists of
minimizing the {\em weighted LS} cost function
\begin{equation} \label{LS-cost}
J_\Delta(\Delta_{1},\dots,\Delta_{{|\Ac|}}) = \sum_{(i,j) \in \Ec(\Ac)} \beta_{j,i} \left ( \widehat{\delta \Delta}_{j \rightarrow i} -  (\Delta_{j} - \Delta_{i}) \right )^2,
\end{equation}
where $\beta_{j,i} = 1/\EE[|z_{j \rightarrow i}|^2]$ is the inverse estimation MSE for the link $j \rightarrow i$.  
Notice that if the estimation errors $z_{j \rightarrow i}$ are Gaussian i.i.d., then  the minimizer of (\ref{LS-cost}) coincides with the joint ML estimator 
based on the observations (\ref{f-estimates}).

Taking the partial derivative with respect to $\Delta_{i}$, and setting it equal to zero, we obtain the set of linear equations
\begin{equation} \label{LS-system}
\sum_{j \in \Ac(i)} (\beta_{i,j} + \beta_{j,i}) \left ( \Delta_{i} - \Delta_{j} \right ) =  \sum_{j \in \Ac(i)} \left (  \beta_{i,j} \widehat{\delta \Delta}_{i \rightarrow j}  -  \beta_{j,i} \widehat{\delta \Delta}_{j \rightarrow i}   \right ), 
\end{equation}
that can be written in matrix form as
$\Lm\Deltam = \uv$, 
where $\Deltam = (\Delta_{1}, \ldots, \Delta_{{|\Ac|}})^\transp$, $\uv$ is the $|\Ac| \times 1$ vector 
with $i$-th element $\sum_{j \in \Ac(i)} \left ( \beta_{i,j} \widehat{\delta \Delta}_{i \rightarrow j} - \beta_{j,i} \widehat{\delta \Delta}_{j \rightarrow i} \right )$, 
and $\Lm$ is the $|\Ac| \times |\Ac|$ matrix with $(i,j)$-th elements
\begin{equation} \label{Lmatrix}
[\Lm]_{i,j} = \left \{ \begin{array}{ll}
\sum_{j \in \Ac(i)} (\beta_{i,j} + \beta_{j,i}) & \;\; \mbox{for} \;\; j = i \\
- (\beta_{i,j} + \beta_{j,i}) & \;\; \mbox{for} \;\; j \neq i \end{array} \right . 
\end{equation}
In general, the actual estimation MSE may not be known. 
Hence, the CRB developed in Appendix \ref{crb-ml} can be used to approximate the coefficients $\beta_{i,j}$, 
motivated by the fact that in the high SNR range  of interest for WLAN applications,  good estimators 
yield MSE essentially identical to the CRB (see results in Appendix \ref{crb-ml}). A simple alternative consists of 
using $\beta_{i,j} = $const. for all $i,j$, that is, using LS instead of weighted LS. 
In particular, by letting $\beta_{i,j} = 1/2$ for all $(i,j) \in \Ec(\Ac)$ we have that $\Lm$ is 
the Laplacian matrix of the connectivity graph of $(\Ac, \Ec(\Ac))$.\footnote{The connectivity graph of the directed graph 
$(\Ac, \Ec(\Ac))$ is the undirected graph with nodes $\Ac$ and edges connecting $i, j$ whenever $(i,j), (j,i) \in \Ec(\Ac)$.
The Laplacian matrix is the $|\Ac| \times |\Ac|$ matrix with elements $-1$ in positions corresponding to $i,j$ whenever these nodes are connected, 
element $|\Ac(i)|$ (the degree of $i$) in the diagonal position corresponding to node $i$ and zero elsewhere.}

It is immediate to see that $\Lm$ has rank $|\Ac| - 1$ and the system of equations (\ref{LS-system}) is under-determined. 
In fact, $\Lm \onev = \zerov$ ($\onev$ being the all-one vector). 
This is to be expected, since only frequency differences can be observed. 
In order to find a solution, we choose a reference anchor AP,  without loss of generality indexed by $1$, and define the frequency 
difference vector $\xiv$ with elements $\xi_i = \Delta_{i} - \Delta_{1}$ for all $i \in \Ac$ with $i \neq 1$. 
With this substitution, we obtain the system of equations $\Lm_1 \xiv = \uv$, where 
$\Lm_1$ is obtained from $\Lm$ by eliminating the first column, yielding the canonical LS solution
$\xiv = ( \Lm_1^\transp \Lm_1)^{-1} \Lm_1^\transp \uv$. 
The resulting correction factor $\Delta^{corr}_{i}$ is equal to $0$ for $i = 1$ (reference anchor AP) and equal to $\xi_i$ 
for $i \neq 1$. 

The TO estimation problem is completely analogous, by defining the corresponding weighted LS cost function 
$\sum_{(j,i) \in \Ec(\Ac)} \gamma_{j,i} \left ( \widehat{\delta \mu}_{j \rightarrow i} -  (\mu_{j} - \mu_{i}) \right )^2$, with
$\gamma_{j,i} = 1/\EE[|w_{j \rightarrow i}|^2]$. The details are omitted for brevity.

\subsection{Numerical Results}
\label{section:simulations-synch}

\begin{figure}
\centering 
\subfigure{\label{fig:rate_v_snr}\includegraphics[width=8cm]{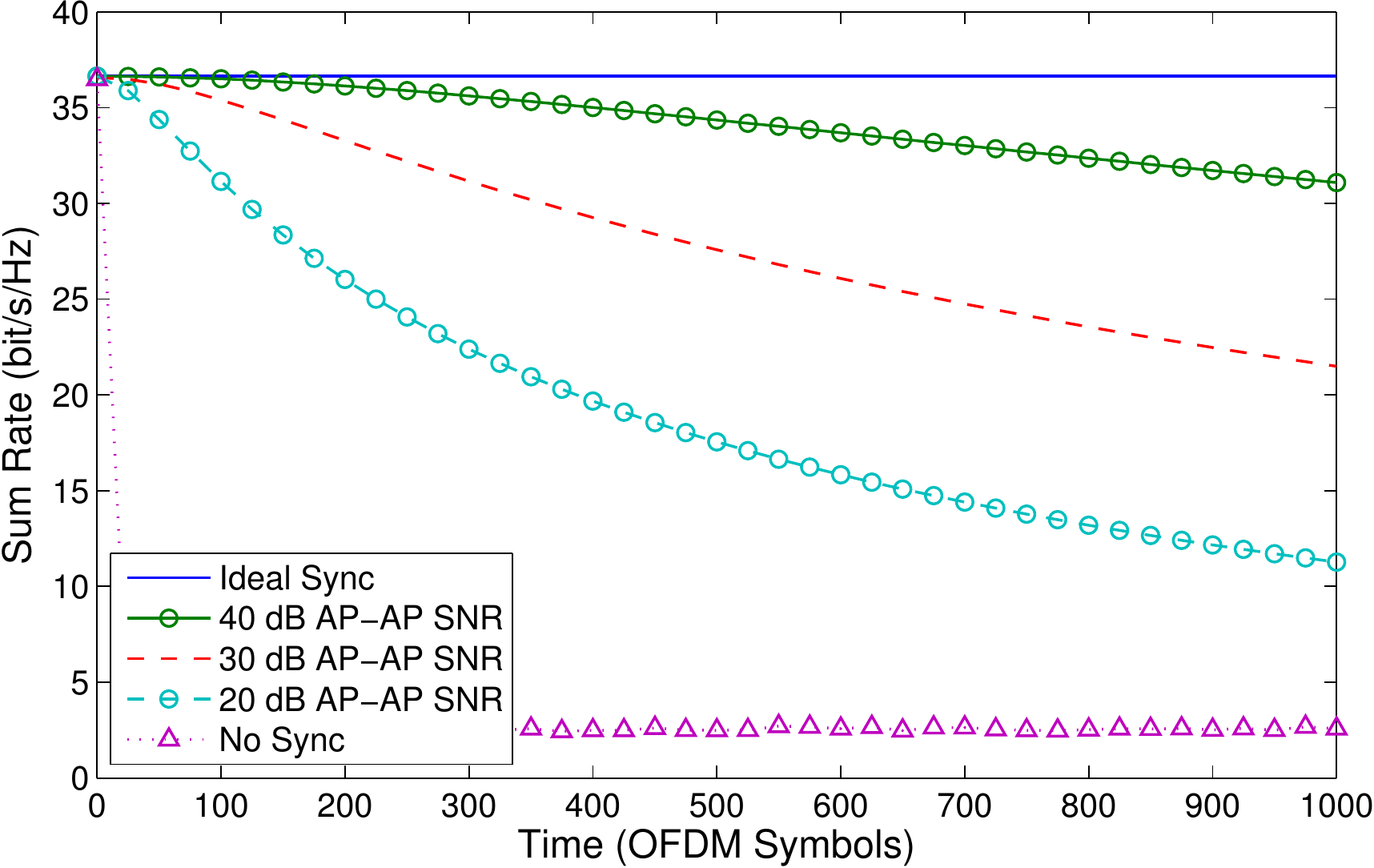}}
\subfigure{\label{fig:rate_v_seq}\includegraphics[width=8cm]{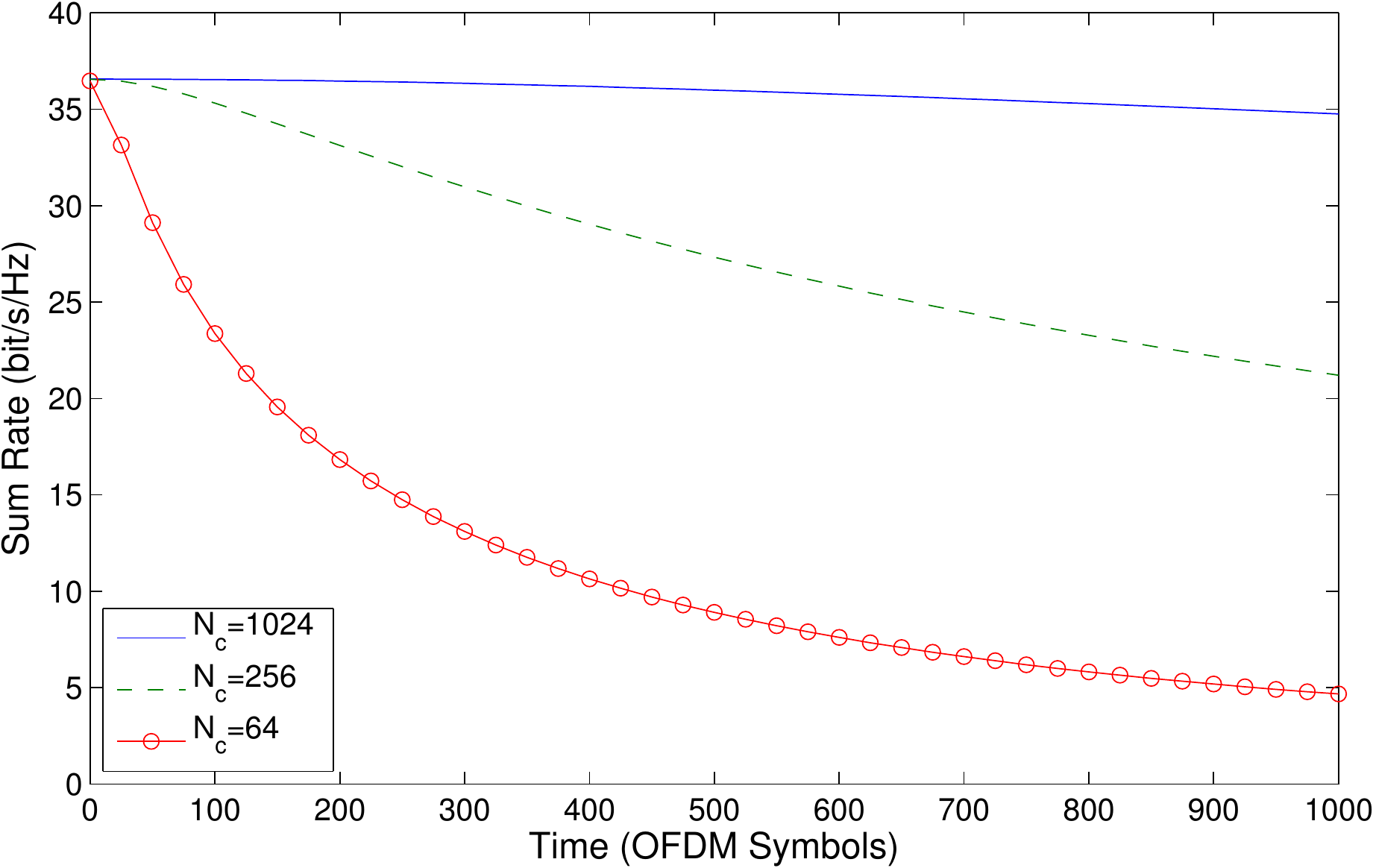}}
\caption{The achievable rates versus time of a $4\times 4$ system after synchronization and compensation, as a function of the OFDM index in the data slot. The left figure assumes $N_c=256$, and the right figure assumes the AP-AP SNR is 30 dB.}
\label{fig:timerates}
\end{figure}

In order to evaluate the impact of TO, CFO and SFO compensation on the performance of distributed MU-MIMO, 
we examine the example of Section \ref{sec:sync_impact} under the proposed synchronization/compensation scheme.  
In this example there are 4 single-antenna APs serving 4 UTs. 
For this small network, we assume that all 4 APs are anchor nodes, in line of sight of each other 
(the channel is formed by a  single path with an unknown  but constant channel coefficient), 
with the same AP-to-AP SNR. We perform timing and frequency joint ML estimation, as derived in Appendix \ref{crb-ml} to estimate 
the offsets for each pair of APs, and then use LS-based synchronization on the fully connected anchor graph. For simplicity, in all the numerical results 
of this paper we have used LS with uniform weights, assuming (conservatively) that the CS does not know the actual estimation MSEs. 

The performance of the TO/CFO estimators between any pair of APs in this example are given in 
Fig.~\ref{fig:mle}-b) in Appendix \ref{crb-ml}, showing 
the MSE vs. SNR relationship of the proposed joint timing and frequency ML estimator along with the associated CRB.  
We notice that even for fairly low SNR this estimator performs very close to the CRB, which decreases inversely proportional to the receiver SNR.

Fig.~\ref{fig:timerates} shows the achievable rate versus the index of the OFDM 
symbol for the case where the channel matrix is perfectly known at time $m = 0$ and, 
as explained before, the ZFBF precoder is calculated at time 0 and kept constant throughout the block.\footnote{Achievable rates are calculated assuming perfect calibration in order to isolate the effects of synchronization.}
While the results of Fig.~\ref{fig:achievablerates} are obtained for free-running oscillators, 
here we assume that at time $m = 0$ the CFO is estimated according to the proposed LS scheme 
and the phase rotation factor (\ref{symbol-rotation}) is applied throughout the sequence of OFDM symbols.  We also assume that the SNR for the AP-to-UT channels is constant at 30 dB. 
For the pilot bursts, we used a structure formed by two repetitions of the same sequence of $N_c/4$ random QPSK time-domain chips,
separated by $N_c/2$ zeros, for a total sequence length of $N_c$ pilot chips (see also the example in Appendix \ref{crb-ml}). 

In Fig.~\ref{fig:timerates} (left), we held the sequence length constant at 
$N_c = 256$ and vary the AP-to-AP SNR.  Consistently with Fig.~\ref{fig:mle}-b), the higher the SNR, the lower the estimation MSE and, as a consequence, 
the residual CFO is smaller and produces a less evident mismatch of the MU-MIMO precoder throughout the data block in  Fig.~\ref{fig:timerates} (left). 
In  Fig.~\ref{fig:timerates} (right), we held the AP-to-AP SNR constant at 30 dB while varying  $N_c$. 
We observe that with $N_c = 1024$ (roughly corresponding to 13 OFDM symbols including the cyclic prefix)  
we can afford to send about 1000 OFDM symbols with small degradation.  
Considering data slots of $\sim 100$ OFDM symbols, we can afford 10 DL MU-MIMO precoded data slots in between each synchronization 
slot. Performance can be improved further by implementing a smoothing filter over time (e.g., Kalman filtering) in order to track
the timing and frequency corrections factors through the sequence of frames instead of re-estimating anew at each frame. 
However, this requires modeling the variations of the clock frequency errors $\epsilon_i$ 
over time (e.g., a Gauss-Markov model) and goes beyond the scope of this paper.

\section{Calibration for TDD Reciprocity}
\label{section:calibration}

In this section we present a novel TDD {\em relative} calibration technique that
generalizes the approach of \cite{argos} to the case of an arbitrary distributed network topology. 
We re-consider the channel model (\ref{bb-model}) assuming that, for clarity of exposition, 
synchronization is perfect, i.e., the residual TO, CFO and SFO are equal to zero. 
Then, we focus on the effects of transmit/receive hardware mismatch on end-to-end channel reciprocity, and devise a scheme able to compensate for such mismatches. 
We remark here that while a centrally clocked MU-MIMO architecture (such as \cite{argos}) needs to perform calibration at a very low rate (e.g., one calibration round every 10 min), due to the inherent stability of the complex scaling introduced by the transmit/receive hardware at each AP, in the case of a 
distributed MU-MIMO architecture as considered in this paper the residual CFO after non-ideal compensation yields a phase error 
that accumulates over the OFDM symbols across the data blocks. 
Hence,  calibration must be repeated together with synchronization, as evidenced by the frame structure of Fig.~\ref{fig:superframe}. 
The calibration slot is formed by pilot symbols arranged on the time-frequency plane induced by OFDM, as commonly implemented for 
channel estimation purposes \cite{seanICC2012,802.11-2012,molisch-book}.
Since the non-reciprocal  elements of the channel (due to the modulation/demodulation hardware)  are smooth over the signal bandwidth 
both in amplitude and in phase, the calibration pilots can be arranged efficiently on the time-frequency plane such that
a large number of mutually orthogonal pilots can be exchanged in a short slot. Furthermore, exploiting the correlation 
between the calibration coefficients at different  subcarriers can be exploited to improve estimation accuracy.
More details on efficient calibration pilot design are
provided in \cite{babis-CTW,babis-CTW-submit} and go beyond the scope of this work. 
Here, for the sake of clarity, we focus on a single subcarrier and drop the subcarrier index $\nu$ from the notation.

In order to motivate the need for TDD reciprocity calibration, let's consider a scheme where the DL channel is learned at the APs side 
through UL pilot symbols sent by the users (with reference to the frame structure of Fig.~\ref{fig:superframe}).
In general, the signal sent from the APs to the UTs over the DL data slot 
at a given subcarrier and OFDM symbol takes on the form
\begin{equation}
\Ym = \Xm \Hm  + \Zm, 
\end{equation}
where the DL channel matrix is given by  $\Hm = \Tm \Bm \widetilde{\Rm}$, 
where $\widetilde{\Rm}  = \diag(\widetilde{R}_1, \ldots, \widetilde{R}_{N_u})$ contains
the complex coefficients introduced by the UT receiver hardware, 
$\Tm  = \diag(T_1, \ldots, T_{N_a})$ contains analogous coefficients 
introduced by the APs transmitter hardware,  and $\Bm$ is the physical channel 
matrix, i.e., the  channel coefficients due solely to the antenna-to-antenna propagation. 

At each given frame,  the UTs that have been polled, send their UL pilots for channel estimation on the UL pilot slot (see Fig.~\ref{fig:superframe}). 
The signal received by the APs is given by 
\begin{equation}
\Ym^{up} =  \Hm^{up}  \widetilde{\Xm}  + \Zm^{up},
\end{equation}
where $\Zm^{up}$ is the UL Gaussian noise vector, 
$\widetilde{\Xm}$ is a $N_u \times N_u$ unitary matrix of frequency domain UL 
pilot symbols, and where $\Hm^{up} = \Rm \Bm \widetilde{\Tm}$ with 
$\widetilde{\Tm}  = \diag(\widetilde{T}_1, \ldots, \widetilde{T}_{N_u})$ containing the UTs transmitter coefficients, 
and  $\Rm  = \diag(R_1, \ldots, R_{N_a})$ containing the APs receiver coefficients. 

Due to TDD (uplink and downlink are in the same frequency band)
and to the reciprocity of the physical propagation channel,  the same $\Bm$ appears in both 
the UL pilot slot and the DL data slot \cite{argos}.

Uplink channel estimation using the UL pilot slot $\Ym^{up}$ can be written as 
$\widehat{\Hm}^{up} = \Ym^{up} \widetilde{\Xm}^\herm = \Hm^{up} + \widetilde{\Zm}^{up}$,
with $\widetilde{\Zm}^{up} = \Zm^{up} \widetilde{\Xm}^\herm$. Neglecting the 
estimation error  $\widetilde{\Zm}^{up}$ for the time being, we notice that if the CS computes the downlink MU-MIMO precoder 
assuming $\widehat{\Hm}^{up}$ as the actual {\em downlink} channel, then this precoder will be mismatched with respect to the true
downlink  channel $\Hm$. For example, with ZFBF precoding, the MU-MIMO precoding matrix obtained from the UL pilots is given
by the two-normalized pseudo-inverse $\Gm = \Lambdam \left [ (\Hm^{up})^\herm \Hm^{up} \right ]^{-1} (\Hm^{up})^\herm$ 
(see Section \ref{sec:sync_impact}).  Clearly, $\Gm \Hm$ is not a diagonal matrix unless $\Hm = \Hm^{up} \Am$ for some invertible 
diagonal matrix $\Am$. 

In order to turn $\Hm$ into a right diagonally scaled version of $\Hm^{up}$, each  AP $i$ multiplies its transmit signal by the coefficient 
$\alpha R_i/T_i$, for some $\alpha\neq 0$ (this is the operation indicated by the $\sf{CAL}$ block in Fig. \ref{fig:UL-DL_fig}). 
Using matrix notation, it can be easily checked that this calibration operation corresponds to 
\begin{equation} 
\alpha \Rm \Tm^{-1} \Hm = \alpha \Rm \Bm \widetilde{\Rm} = \alpha \Hm^{up} \widetilde{\Tm}^{-1} \widetilde{\Rm}, 
\end{equation}
where the diagonal matrix $\Am =  \alpha \widetilde{\Tm}^{-1} \widetilde{\Rm}$ contains the scalar amplitude and phase shifts 
that are estimated at each UT receiver as part of their DL channels.
It is therefore apparent that the calibration protocol consists of efficiently estimating the TDD 
calibration matrix $\alpha \Rm \Tm^{-1}$, defined up to an arbitrary multiplicative non-zero  
factor $\alpha$. As for the synchronization protocol of Section \ref{section:synchronization}, 
also in this case we devise a scheme that involves only the exchange of calibration pilots between APs, i.e., our scheme does not
assume the participation of the UTs, which may be legacy devices not designed to handle distributed MU-MIMO.

During the calibration slot, pilot symbols are exchanged by the APs over a connected 
spanning subgraph  $(\Tc, \Fc)$ of the network graph ($\Tc, \Ec)$ where the subset of links $\Fc$ is such that if 
$(i,j) \in \Fc$ then also $(j,i) \in \Fc$. 
The optimization of the subgraph $(\Tc, \Fc)$  is an interesting topic for future work. 
 
Let $(i,j) \in \Fc$. Then, after a calibration slot,   AP $i$ gathers the observation
\begin{equation}  \label{sucamillo}
Y_{j \to i} = T_j \,B_{j \to i} \, R_i + Z_{j \to i}, 
\end{equation}
and AP $j$ gathers the observation
\begin{equation} 
Y_{i \to j} = T_i \,B_{i \to j} \, R_j + Z_{i \to j}
\end{equation}
where we have assumed without loss of generality that the calibration pilot transmitted by the APs is equal to one.
We can write
\begin{eqnarray}
\left [ \begin{array}{c} Y_{j\to i} \\ Y_{i\to j} \end{array} \right ] & = & 
\left [ \begin{array}{c} T_j\, R_i   \\ T_i \, R_j \end{array} \right ] B_{i\to
j} + 
\left [ \begin{array}{c} Z_{j\to i} \\ Z_{i\to j} \end{array} \right ] \nonumber
\\
& = & \left [ \begin{array}{c} c_i  \\ c_j \end{array} \right ]  \rho_{i,j} + 
\left [ \begin{array}{c} Z_{j\to i} \\ Z_{i\to j} \end{array} \right ],
\label{observationy}
\end{eqnarray}
owing to the fact that, by the physical channel reciprocity, $B_{i\to j} = B_{j \to i}$, and defining
$\rho_{i,j}  = \rho_{j,i} = T_i \, T_j \, B_{i\to j}$ and $c_i = R_i/T_i$. 
Our goal is to estimate the relative calibration coefficients $c_i$ for $i =
1,\ldots, {N_a}$, up to a common multiplicative 
non-zero constant $\alpha$. Without loss of generality we assume that $\{c_i\}$ is a set
of non-zero bounded complex numbers.\footnote{If $c_i=0$ or $1/c_i=0$, the
$i$-th node can be omitted as it is a ``non-communicating'' node.} 

If the observations $Y_{j \to i}, Y_{i \to j}$ were noiseless, we would have the equality
$c_j Y_{j \to i} = c_i Y_{i \to j}$ for all $(i,j) \in \Fc_u$, where $\Fc_u$ is the set of undirected edges corresponding to $\Fc$.
Hence, in the presence of observation noise it makes sense to define the LS cost function
\begin{equation} \label{LS-cal}
J_{cal} (c_1, c_2, \ldots, c_{N_a})  = \sum_{(i,j) \in \Fc_u} \left | c_j Y_{j\to i}  - 
c_i Y_{i\to j} \right |^2.
\end{equation}
Letting $\cv = (c_1, c_2, \ldots, c_{N_a})$, notice that the LS objective function satisfies
the scaling property $J_{cal} (\alpha \cv) = |\alpha|^2 J_{cal} (\cv)$.
In fact, given a vector of calibration coefficients $\cv$, any scaled vector $\alpha \cv$ for $\alpha \neq 0$ provides
equally good calibration coefficients.  In order to eliminate this ambiguity and at the same time exclude the all-zero solution 
in the minimization of  (\ref{LS-cal}), we impose the constraint $\|\cv\|^2 = 1$. 
The Lagrangian function of the constrained LS problem is given by 
\begin{equation} \label{lagrangian}
F_{cal}(\cv,\lambda) = J_{cal}(\cv) - \lambda (\|\cv\|^2 - 1).
\end{equation}
where $\lambda$ is the Lagrangian multiplier.
Differentiating $F_{cal}$ with respect to $c_i^*$, treating $c_i$ and $c_i^*$  as if they were independent
variables \cite{derivatives-book}, and then setting the partial derivatives to zero,  we obtain
\begin{equation}
\frac{\partial}{\partial c_i^*} F_{cal}(c_1,c_2,\ldots, c_{N_a},\lambda) = \sum_{j : (i,j) \in \Fc_u} \left ( c_i |Y_{i\to j}|^2 - c_j Y_{i \to j}^* Y_{j \to i} \right )-\lambda c_i.
\end{equation}
In matrix form, we obtain $\Am \cv = \lambda\cv$, where $\Am$ is the ${N_a} \times {N_a}$ matrix with 
$(i,j)$-th elements 
\begin{equation}
[\Am]_{i,j}  = \left \{ \begin{array}{ll}
\sum_{j : (i,j) \in \Fc_u} |Y_{i\to j}|^2 & \mbox{for} \;\; j = i \\
- Y_{i \to j}^* Y_{j \to i} & \mbox{for} \;\; j \neq i, \;\; (i,j) \in \Fc_u \\
0 & \mbox{for}  j \neq i, \;\; (i,j) \notin \Fc_u \end{array} \right . 
\end{equation}
Finally, we notice that the sought constrained LS solution $\cv$ is a unit-norm eigenvector 
associated to the eigenvalue of $\Am$ with the smallest magnitude. In fact, this yields the direction in the domain $\CC^{N_a}$ of 
$J_{cal}(\cv)$ with slowest growth, such that the value of $J_{cal}(\cv)$ is minimized at the intersection of the unit circle
$\|\cv\|^2 = 1$ and such eigenspace of $\Am$. 

It is interesting to notice the difference between the proposed method and the scheme in \cite{argos}. 
When the graph $(\Tc, \Fc)$ is a star with (say) AP 1 at the center, \cite{argos} proposes to
estimate the calibration coefficients as $c_j = \frac{Y_{1 \to j}}{Y_{j \to 1}} c_1$ for some arbitrary unit-magnitude $c_1$. 
This is equivalent to minimize the objective function (\ref{LS-cal}) subject to the constraint $|c_1| = 1$.
The problem with this method is that the $c_j$'s are given by the ratio of two noisy pilot symbols. In a distributed topology when
SNR to the center star may not be large, this yields an ill-behaved estimation where some coefficients may be near zero and 
others  be very large. When using these coefficients in the MU-MIMO beamforming scheme, subject to a per-AP power constraint, 
the system performance incurs significant degradation unless all APs have very good SNR to the 
star center. 

A final remark about scalability is in order. 
For general network graphs, the computational complexity of LS for both synchronization and TDD calibration 
is polynomial in the number of APs. For certain network graphs this complexity can be linear (e.g., in the star topology of the 
Argos scheme \cite{argos}). In any case, the computation involved in solving the LS estimation problems is easily affordable 
for network of practical size (up to a few tens of APs). It is also interesting to remark that the data exchange over the wired backhaul 
network required by the proposed protocol is also easily affordable. For example, 
suppose that the timing and frequency measurements (real numbers) and the calibration 
pilot symbol (complex numbers) are represented with 16 bits per real coefficient.  
This requires 64 bits per frame per AP (roughly speaking).  For frames of 10 ms, this is equivalent to 6.4 kbit/s per AP of protocol data 
overhead. For a 1 Gb Ethernet backbone, as it would be meaningful for a distributed MU-MIMO system, 
this overhead is less than 5 orders of magnitude less than the backhaul capacity. Therefore, 
a system with 100 APs would consume less than 0.1\% of the backhaul capacity. 

\subsection{Numerical Results}
\label{section:simulations}

\begin{figure}
\centerline{\includegraphics[width=10cm]{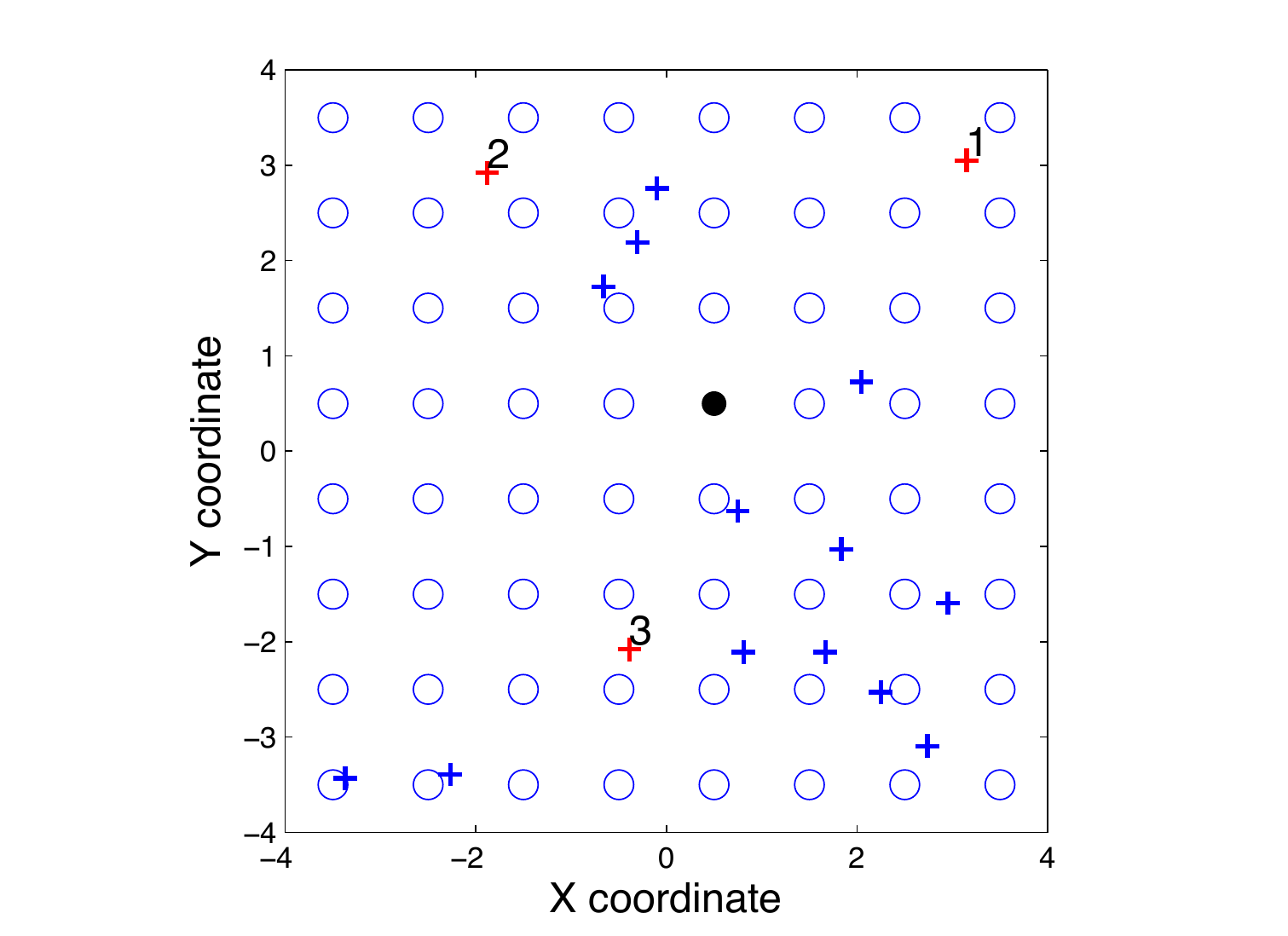}}
\caption{Sample topology involving APs (depicted by ``o'') arranged on an
$8\times8$ grid, and $16$ UTs (depicted by ``$+$''). The reference antenna used for Argos 
calibration is denoted by a $\bullet$.}
\label{fig:simulated_topology}
\end{figure}

We provide a simulation-based comparison between the calibration scheme of \cite{argos} and our scheme. 
For convenience, we refer to the former as  ``Argos-Calibration''  and 
to the latter as ``LS-Calibration''. 
We consider a scenario comprising 64 single antenna APs distributed over a regularly spaced $8 \times 8$ squared 
grid, as shown in Fig.~\ref{fig:simulated_topology}. 
The system serves 16 UTs simultaneously, using distributed MU-MIMO.  The UTs are independently and randomly located with 
uniform probability over the square.  The users' achievable rates are calculated by Monte Carlo simulation, randomizing over several realizations of 
the users' locations.  One such realization is shown in Fig.~\ref{fig:simulated_topology}. 
The distance between the two most distant nodes is 100m. Hence, the minimum distance in the regular grid arrangement between 
APs is equal to $100/7/\sqrt{2} \approx 10.1$m. 

In order to isolate the impact of the TDD reciprocity calibration on the performance of MU-MIMO precoding, 
we assume perfect synchronization. The UL and DL channels are given by 
\begin{align}
Y_{i}^{up}[m,\nu] &= R_i   \sum_{j=1}^{N_{\rm U}} \pi_{i,j} \, B_{i,j}[m,\nu]  \widetilde{T}_j
\, X_j^{up}[m,\nu] + Z^{up}_i[m,\nu]  \label{cl-to-ap-link}\\
Y_{j}[m,\nu] &= \widetilde{R}_j \sum_{i=1}^{N_{\rm A}} \pi_{i,j}\,  B_{i,j}[m,\nu]        
  T_i  \, X_i[m,\nu] +Z_j[m,\nu],  \label{ap-to-client-link}\
\end{align}
where $Z_j[m,\nu]$ and $Z^{up}_j[m,\nu]$ are i.i.d. $\sim \Cc\Nc(0,N_0)$ additive Gaussian noise samples, 
the real-nonnegative scalar $\pi_{i,j}$ denotes the large-scale path gain between AP $i$ and UT $j$ and
we assume i.i.d. small-scale fading $B_{i,j}[m, \nu] \sim \Cc\Nc(0,1)$. 

The large-scale path gains between AP-AP or AP-UT are both based 
on the WINNER model \cite{winner-model}, where the pathloss ($\rm PL$) is given in dB as a function of distance 
($d$, in meters), carrier frequency ($f_0$, in GHz), and log-normal shadowing ($\rho_{\rm {dB}}$ with variance 
$\sigma^2_{\rm {dB}}$) as:
\begin{equation} \label{eq:PL-winner}
\rm{PL} (d)= A \log_{10}(d)+ B + C\log_{10}(f_0 / 5)+ \rho_{\rm {dB}}, \;\;\; 3 \leq d \leq 100,
\end{equation}
where the parameters $A$, $B$ $C$ and $\sigma^2_{\rm{dB}}$ 
are scenario-dependent. We consider an indoor office scenario\footnote{These values correspond to the  
so-called A1 Indoor Office scenario with single, light walls in every path and where all of the UTs and 
APs are located on the same floor.} 
where $A = 18.7, B=46.8, C=20, \sigma^2_{\rm {dB}}=9$ when in line of sight, 
otherwise $A = 36.8, B = 43.8, C=20, \sigma^2_{\rm {dB}}=16$ when not in line of sight. 
For distances $d < 3$m, we conservatively extend the model by setting ${\rm PL}(d)={\rm PL}(3)$. 
This is justified by the fact that the extremely high receive powers associated with shorter distances do not lead to 
higher link capacities, due to practical constraints on the modulation order as well as the receiver hardware 
(gain control and ADC range). The line of sight probability is given by a Bernoulli distribution with parameter 
$p_{los}$, which depends on distance as follows:
\begin{align*}
  p_{los}= 
    \begin{cases}
      1 & d \leq 2.5m \\
      1 - 0.9(1-(1.24-0.6\log_{10}(d))^3)^{1/3} & \text{else.}
    \end{cases}
\end{align*}
The large-scale path gain between AP $i$ and UT $j$ at distance $d_{i,j}$ 
is given by  $\pi_{i,j} = 10^{-(\rm{PL}(d_{i,j})/20)}$. 

In the TDD calibration protocols, pilot signals are transmitted between APs.
In particular, when AP $j$ transmits a pilot symbol  $X_j[m,\nu] $,  
AP $i$ receives 
\begin{equation}
Y_{j \to i}[m,\nu] \!=\! R_i \pi_{j \to i}  B_{j \to i}[m,\nu]  T_j X_j[m,\nu]
\!+\! Z_{j \to i}[m,\nu]. 
\label{ap-to-ap-link}
\end{equation}
$\pi_{j\to i} =\pi_{i\to j}$ denotes the large-scale path gain between APs $i$ and $j$ and follows the
same model used for $\pi_{i,j}$ . The small scale fading coefficients
$B_{i \to j}[m, \nu]$ have the same statistics as $B_{i,j}[m,\nu]$ and we have 
$B_{i \to j}[m, \nu] = B_{j \to i}[m, \nu]$ thanks to the TDD reciprocity.

The hardware-induced non-reciprocal coefficients
$\{R_i, \,T_i\}$ and $\{\widetilde{R}_j, \, \widetilde{T}_j\}$ 
are modeled as i.i.d.  complex random variables with uniformly distributed phase over $[-\pi,\pi]$ and
uniformly distributed magnitude in  $[1 - \varrho, 1 + \varrho]$ with 
$\varrho$ chosen such that the standard deviation of the squared-magnitudes 
is equal to 0.1. 

\begin{figure}
\centering
\subfigure{\includegraphics[width=8cm]{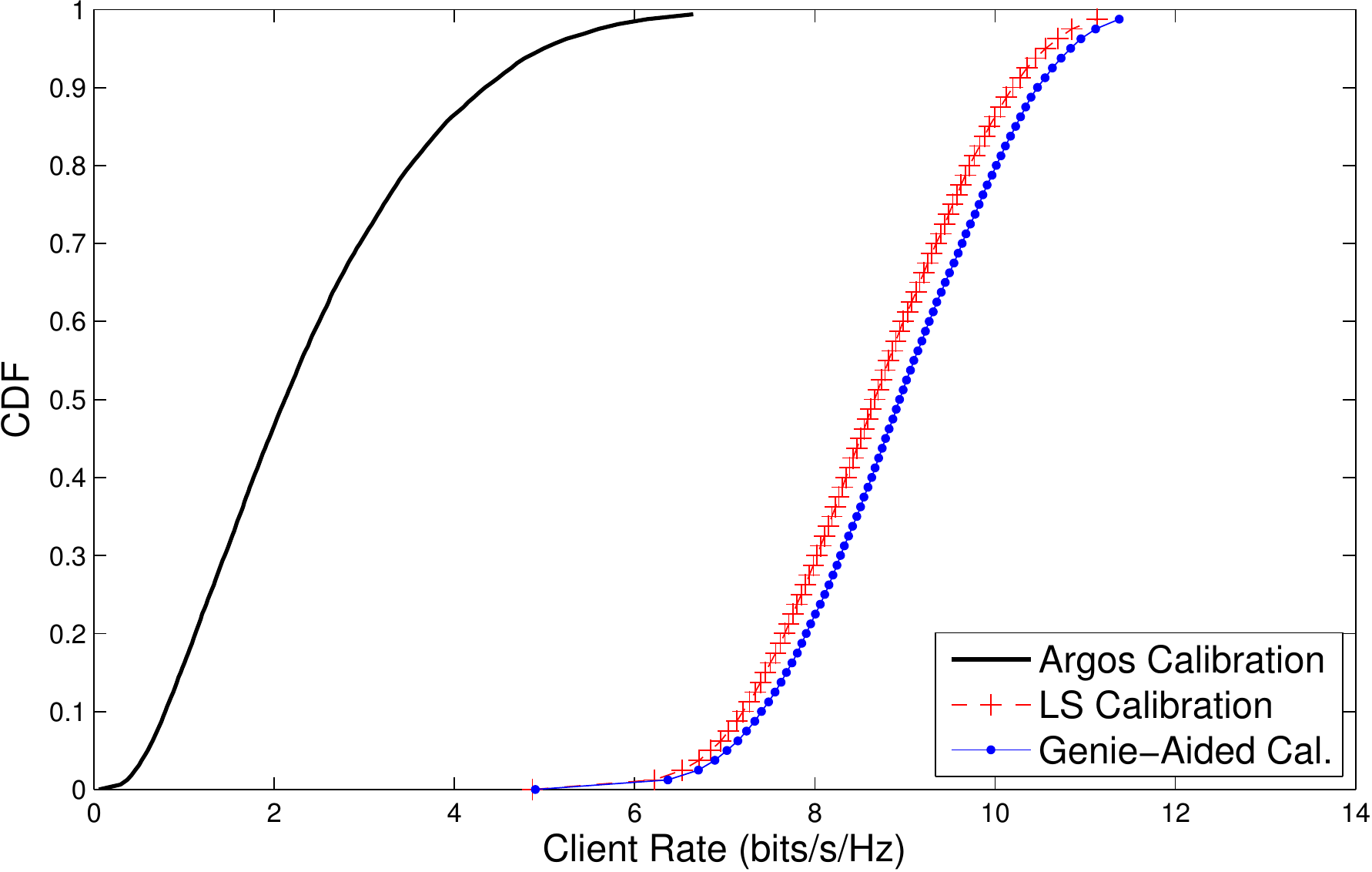}}
\subfigure{\includegraphics[width=8cm]{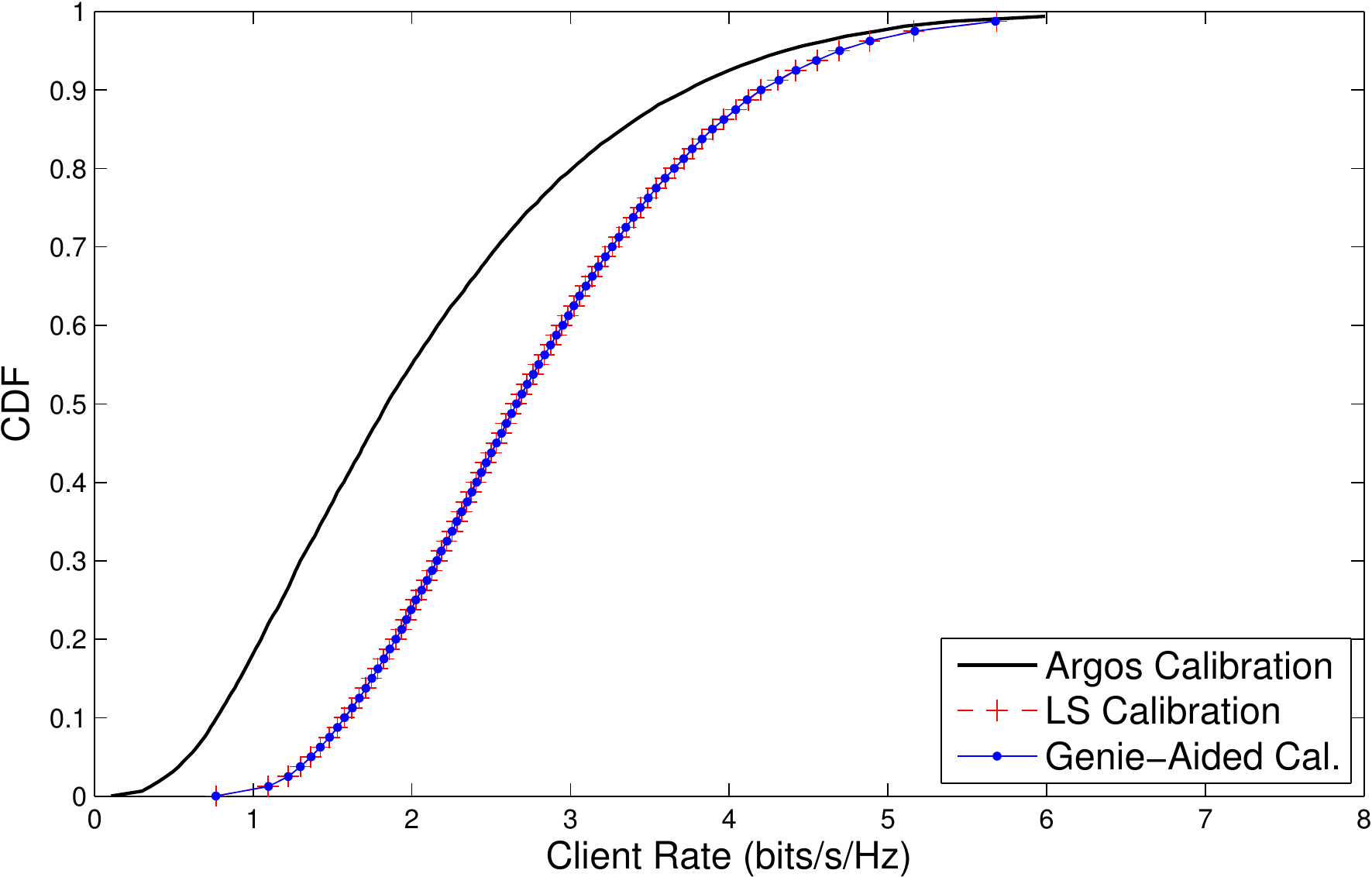}}
\caption{The CDFs of the user rates, across the user locations and the calibration estimates.  The left figure displays the results for ZFBF, while the right 
displays the results for conjugate beamforming.}
\label{fig:all_users}
\end{figure}

\begin{figure}
\centering
\subfigure{\includegraphics[width=8cm]{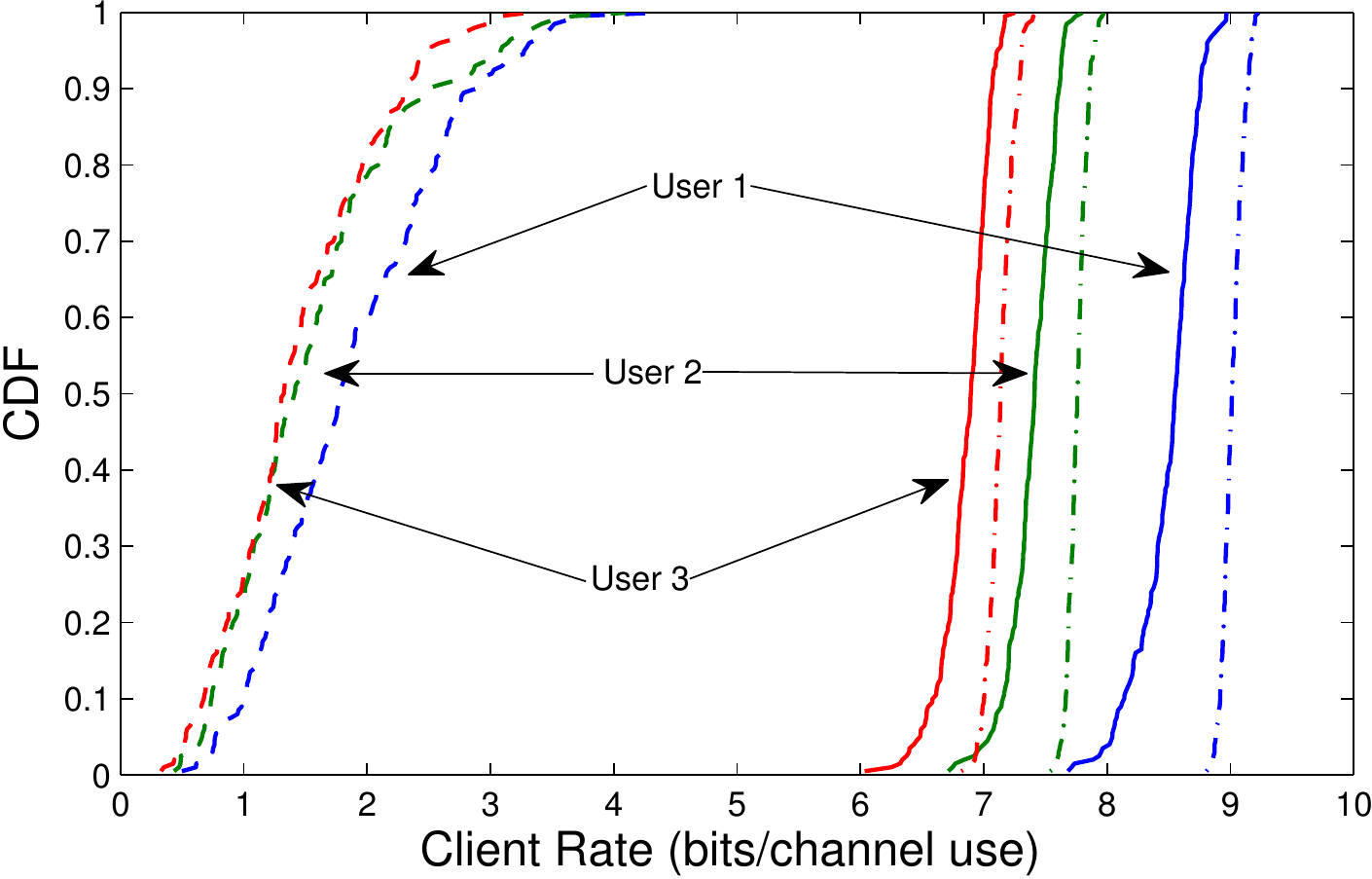}}
\subfigure{\includegraphics[width=8cm]{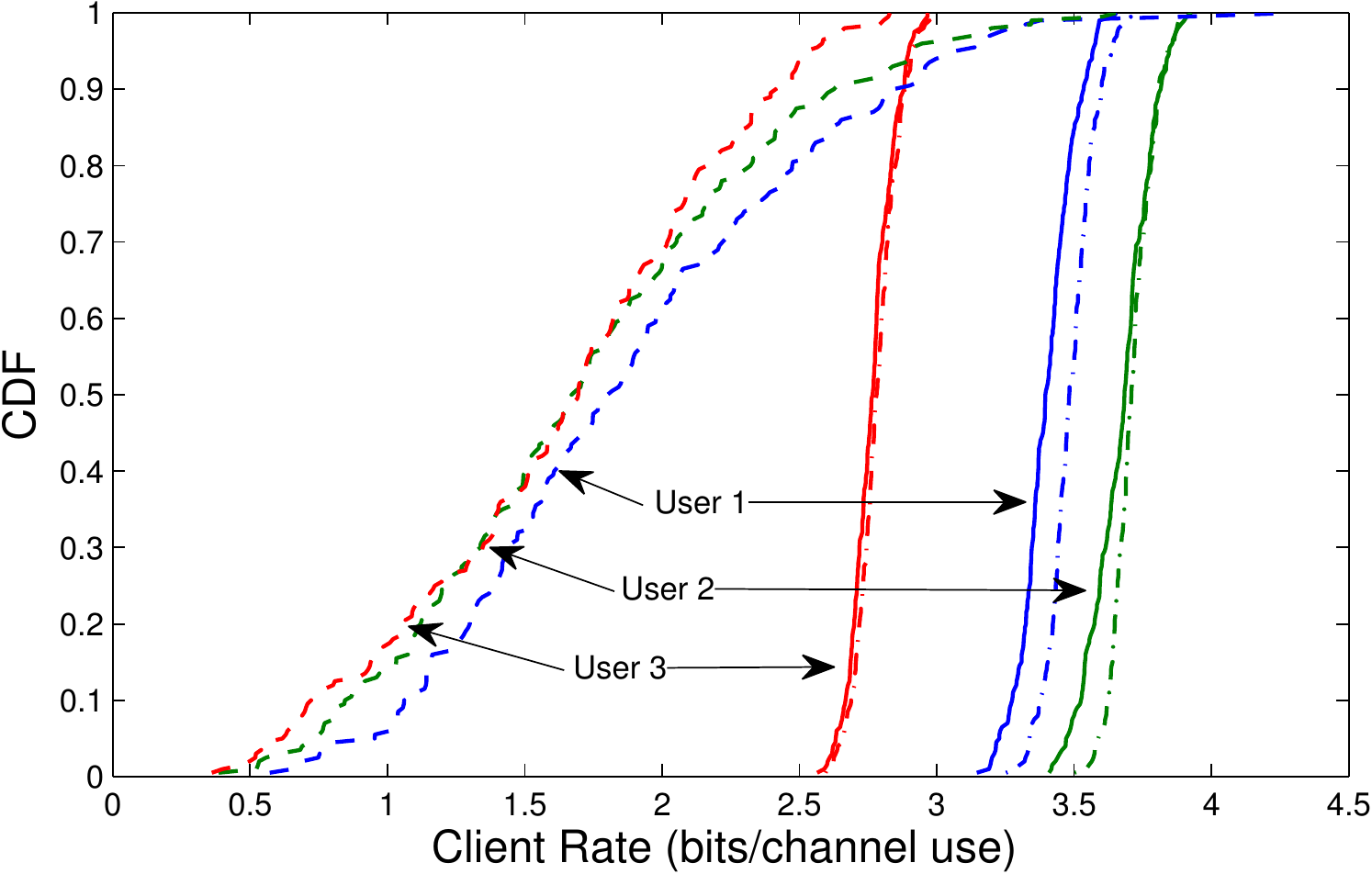}}
\caption{Rate CDFs using ZFBF (left) and conjugate beamforming (right), with
Argos-Calibration (dashed), LS-Calibration (solid) and genie-aided calibration (dash-dot),  
for the UTs indicated by ``1", ``2'' and ``3" in Fig.~~\ref{fig:simulated_topology}.}
\label{fig:zf_cal_vs_genie}
\end{figure}

Fig.~\ref{fig:all_users} shows the performance in terms of achievable rate cumulative distribution function (CDF), obtained by generating
independent realizations of the channels, of the calibration estimation, and of the UT positions. In these results we have assumed (for simplicity)
that all UTs are served with equal power per user, and we have scaled
the transmit signals by a common scaling factor such that the transmit power at each antenna does not exceed the per antenna 
power constraint of 90 dB.\footnote{Backing off the transmit power in order to meet the per-AP power constraint
is suboptimal. The optimal MU-MIMO zero-forcing precoding with per-antenna power constraint is discussed in detail in \cite{huh-power-opt}, and its optimization requires non-uniform power allocation per DL data stream and can be obtained using the theory of generalized inverses and 
convex relaxation. Here, we chose to use a more practical suboptimal approach for the sake of simplicity.}
The reference antenna location used in Argos-Calibration is also indicated in Fig.~\ref{fig:simulated_topology}. 
LS-Calibration is run using a fully connected graph. Genie-aided calibration uses the true values of 
$T_i$'s and $R_i$'s to calculate the calibration coefficients. It should be noted that the pilot overhead for the LS and Argos schemes are identical.

The performance of different calibration schemes is affected by the SNR value between antennas. 
As reported in \cite{argos}, Argos-Calibration requires a careful placement of the reference antenna 
such that it has high enough SNR to all the other antennas. 
Argos, which was designed for co-located antennas, does not perform well in the distributed case due to the large pathlosses between the reference antenna and the more distant antennas. Since LS-Calibration does not depend on a single reference antenna, 
its performance is much less sensitive to the quality of a single AP-to-AP channel, and it achieves essentially genie-aided performance.

In Fig.~\ref{fig:zf_cal_vs_genie}, we present the rates of three particular users (labeled 1, 2 and 3 in Fig.~\ref{fig:simulated_topology}) for the particular realization of the user locations show in Fig.~ \ref{fig:simulated_topology}.

\section{Conclusions} \label{section:conclusion}

In this work we presented scalable solutions for two of the main implementation hurdles of distributed MU-MIMO downlink.  
First, we motivated the need for timing and frequency synchronization, and detailed the performance degradation 
in the case of uncompensated frequency offsets between the jointly precoded APs. 
Next, we outlined a network-wide synchronization procedure based on a pilot burst exchange protocol, 
local timing and frequency offset estimation, and a centralized constrained LS solution for the overall timing and frequency 
correction factors.  Finally, we introduced a novel method for TDD reciprocity calibration so as to enable DL MU-MIMO based on UL training, even though
the APs are equipped with a standard, non-reciprocal,  RF front-end.  The proposed calibration protocol method, unlike its predecessors, 
yields near-ideal performance with distributed 
node deployment. Notice that  both synchronization and calibration involve only the APs, and do not assume any UT collaboration. 
Therefore, the proposed schemes are suited to work with ``legacy'' user equipment, not explicitly designed for this purpose.
Also, all the proposed schemes are immediately applicable to APs equipped with multiple antennas, where the groups of antennas belonging to the same AP
are driven by a common clock and therefore need only to be calibrated but not mutually synchronized. Hierarchical calibration in multiple steps
is naturally suited to this purpose, as evidenced in \cite{calibration-ita2013,babis-CTW}. 
It is also interesting to notice that the proposed synchronization scheme calculates the timing and frequency correction factors relative to 
some particular anchor AP (denoted by AP 1 in Section \ref{section:synchronization}). Hence, a side benefit of our scheme is that 
network-wide timing and frequency stability can be achieved by having a single node (denoted as AP 1) equipped with a very stable oscillator, without the need to have a high-SNR connection from each node to this particular AP.
Finally, extensive simulation results and experimental evidence provided by a software radio implementation
\cite{mobicom2012,airsync-ton}, we conclude that the proposed schemes effectively enable the implementation of 
distributed MU-MIMO network architectures, and can turn a cluster of small cell APs into a large distributed cooperative 
antenna system. 

\appendices

\section{Timing and Frequency Joint ML Estimation}
\label{crb-ml}

We consider joint TO and CFO pilot-aided estimation for a transmitter-receiver pair, where the channel
is multipath time-invariant with additive white Gaussian noise, with known path delays and unknown
path coefficients.  
We denote by  $\delta\tau$ and by $\delta f$ the timing and carrier frequency differences 
between the transmitter and the receiver.

The received time-domain baseband signal corresponding to a pilot burst transmission is given by 
\begin{equation}  \label{received-pilot-burst}
y(t) = \left ( \sum_{p=0}^{P-1} h_p s(t - \delta \tau - \rho_p) \right )
e^{j2\pi \delta f t} + z(t) 
\end{equation} 
where the multipath channel impulse response is $h(t) =  \sum_{p=0}^{P-1} h_p \delta(t - \rho_p)$, 
and where the pilot signal is given by 
\begin{equation}
s(t) = \frac{1}{\sqrt{T_s}} \sum_{n=0}^{N_c-1} s_n \Pi(t/T_s - n),
\end{equation}
for the sequence of time-domain chip symbols $\sv = (s_1,\ldots, s_{N_c})$.
In (\ref{received-pilot-burst}), $z(t)$ denotes the  complex circularly-symmetric Gaussian white noise, with autocorrelation 
function $\EE[z(t) z^*(t - \tau)] = N_0 \delta(\tau)$. 
Letting $\nabla(t) = (\Pi \otimes \Pi)(t)$ and assuming that the receiver 
performs chip matched filtering and sampling with one sample per chip, we obtain the received discrete-time observation
\begin{equation}
y[m] = \left ( \sum_{p=0}^{P-1} h_p \sum_{n=0}^{N_c-1} s_n \nabla \left ( m - n -
\frac{\delta \tau + \rho_p}{T_s} \right ) \right ) e^{j2\pi \delta f T_s m} + z[m]
\end{equation}
where $z[m] \sim \Cc\Nc(0,N_0)$ is a discrete-time complex circularly symmetric i.i.d. Gaussian process.
As said in Section \ref{section:architecture}, the receiver has a coarse knowledge of the frame timing 
such that it expects to find the pilot burst  in a given interval including the guard times. 
The receiver collects $M$ samples such that the sampled interval of duration 
$MT_s$ contains the pilot burst. We collect the $M$ received samples into an $M \times 1$ 
vector $\yv = (y[0], \ldots, y[M-1])^\transp$. We define 
$\delta \xi = \delta f T_s$,\footnote{This is related to $\Delta_i,\Delta_j$ as defined in (\ref{norm-offset}) by 
$\delta \xi = \frac{\Delta_i - \Delta_j}{N + L}$,  when $i$ is the transmitter and $j$ the receiver.} 
and $\delta \mu = \delta \tau / T_s$, consistently with the normalized TO defined in  (\ref{norm-offset}). 
We define the diagonal matrix 
\[ \Xim(\delta\xi) = \diag \left ( 1, e^{j2\pi \delta \xi}, e^{j4 \pi \delta \xi}, \ldots, e^{j2\pi \delta \xi  (M-1)} \right ), \]
the $M \times (M - N_c + 1)$ convolution (tall Toeplitz) matrix
\[ \Sm = \left [ \begin{array}{cccccc}
s_0  &  0   &  \cdots  &     &      &   0        \\
s_1 & s_0 &              &    &       &  \vdots \\
\vdots &    & \ddots   &     &      &             \\
s_{N_c-1} &               &      &     &  &  \vdots \\
0               & \ddots   &      &     &    &  s_0   \\
\vdots       &               &      &     &   & \vdots \\
0              & \cdots     &      &     &    & s_{N_c-1} \end{array} \right ],
\]
and the matrix $\mathbf{\nabla}(\delta\mu)$ of dimension $(M - N_c + 1) \times P$ 
whose $p$-th column (for $p = 0,\ldots, P-1$) has entries
\[ \nabla \left (\ell  - \delta \mu  - \rho_p/T_s \right ), \;\;\; \mbox{for}
\;\; \ell = 0, \ldots, M - N_c. \]
In this way, the discrete-time sampled version of the convolution $(\nabla \otimes h)(t)$ is 
given by the product $\mathbf{\nabla}(\delta\mu) \hv$, where $\hv = (h_0,\ldots, h_{P-1})^\transp$ is the vector of 
path coefficients.  Eventually we arrive at the vector observation model
\begin{equation} 
\yv =  \Gammam(\delta\xi, \delta\mu) \hv + \zv,
\end{equation}
where $\Gammam(\delta\xi, \delta\mu)  = \Xim(\delta \xi) \Sm \mathbf{\nabla}(\delta \mu)$
and  $\zv = (z[0],\ldots, z[M-1])^\transp$.
 
The joint ML estimator for $\delta \xi, \delta\mu$ and $\hv$, assuming $N_0$ and  $\{\rho_p\}$ known, 
is obtained by maximizing the Likelihood Function
\begin{equation}  \label{likelihood-function}
\Lambda (\delta \xi, \delta\mu, \hv) = \frac{1}{(\pi N_0)^M} \exp \left ( -
\frac{1}{N_0} \left \| \yv  - \Gammam(\delta \xi, \delta \mu) \hv \right \|^2 \right ), 
\end{equation}
or equivalently by minimizing the square distance  $\left \| \yv  - \Gammam(\delta \xi, \delta \mu) \hv \right \|^2$. 
For given $\delta\xi$ and $\delta \mu$, the minimization with respect to $\hv$ yields the classical LS solution
\[ \widehat{\hv} =   \left ( \Gammam^\herm (\delta \xi, \delta \mu) \Gammam(\delta \xi, \delta \mu)  \right )^{-1} \Gammam^\herm(\delta \xi, \delta \mu)\yv. \]
Replacing this into the objective function, after some simple algebra, 
we find that the joint ML estimator is obtained by maximizing with respect to $\delta\xi$ and $\delta\mu$ the quadratic form
\begin{equation}
F(\delta\xi, \delta\mu) =  \yv^\herm \Gammam(\delta \xi, \delta \mu) \left ( \Gammam^\herm (\delta \xi, \delta \mu) \Gammam(\delta \xi, \delta \mu)  \right )^{-1} \Gammam^\herm(\delta \xi, \delta \mu) \yv.
\end{equation}
This can be interpreted as maximizing the energy of the projection $\yv$ onto the column space of the signal matrix 
$\Gammam(\delta \xi, \delta \mu)$.  Eventually, the joint ML estimator is obtained by searching over a two-dimensional 
sufficiently fine grid  of points with respect to the variables $\delta\xi$ and $\delta\mu$. 

Next, we derive the CRB for this estimation problem. 
It is well-known that when the likelihood function is a multivariate Gaussian pdf
$\Cc\Nc( \mv(\thetav) , \Sigmam)$ with the dependence on the parameters in the
mean, then  the Fisher information matrix $\Jm(\thetav)$ has $(a,b)$ elements 
\cite{kay-estimation}
\[ J_{a,b}(\thetav) = 2\text{Re}\left \{ \left(\frac{\partial \mv(\thetav)}
{\partial \theta_a}
\right )^\herm \Sigmam^{-1} \frac{\partial \mv(\thetav)}{\partial \theta_b} \right\}. \]
In our case, we have $\Sigmam = N_0 \Id$, $\mv (\thetav) =  \Xim(\delta \xi) \Sm \nabla(\delta \mu) \hv$, and the vector of parameters $\thetav$
has elements
\[ \theta_1 = \delta \xi, \theta_2 = \delta\mu, \theta_{2p+3} = \text{Re}\{h_p\}, \theta_{2p+4} = \text{Im}\{h_p\}, \; p = 0,\ldots, P-1. \]
Define $\Dm = \diag ( 0, 1, \ldots, M-1 )$ and let $\ev_p$ denote $P\times 1$ vector with all zeros and a single 1 in position $p$, for 
$p = 1,\ldots, P$.  Also let $\mathbf{\nabla}'(\delta\mu)$  denote the derivative of $\mathbf{\nabla}(\delta\mu)$ 
with respect to $\delta\mu$,  whose $p$-th column has only two non-zero elements equal to\footnote{This is the only step where 
we make explicit use of the fact that $\nabla(t)$ is a triangular waveform.}
$-1$ at component $\ell =  \left \lfloor \delta \mu + \rho_p/T_s \right \rfloor$ and 
equal to $+1$ at component $\ell = \left \lfloor \delta \mu + \rho_p/T_s \right \rfloor+1$. Then,  we can write
\begin{eqnarray} 
\frac{\partial \mv(\thetav)}{\partial \delta\xi} & = & j2\pi \Dm  \Xim(\delta \xi) \Sm\nabla(\delta\mu) \hv, \label{partial-xi} \\
\frac{\partial \mv(\thetav)}{\partial \delta\mu} & = & \Xim(\delta\xi)\Sm \nabla'(\delta\mu) \hv, \label{partial-mu} \\
\frac{\partial \mv(\thetav)}{\partial \Re\{ h_p\}} & = & \Xim(\delta\xi)\Sm \nabla(\delta\mu) \ev_p, \label{partial-realc} \\
\frac{\partial \mv(\thetav)}{\partial \Im\{ h_p\}} & = & j \Xim(\delta\xi) \Sm \nabla(\delta\mu)\ev_p. \label{partial-imagc}
\end{eqnarray}
Defining the  $M \times 2(P+1)$ matrix $\Km = \left [\frac{\partial \mv(\thetav)} {\partial \theta_1}, \frac{\partial \mv(\thetav)} {\partial \theta_2}, \ldots, \frac{\partial \mv(\thetav)} {\partial \theta_{2(P+1)}}\right ]$, the Fisher information matrix is obtained as
\begin{equation}
\Jm(\thetav) = \frac{2}{N_0} \text{Re} \left\{ \Km^\herm \Km \right \}.
\end{equation}
Eventually, the CRB for $\delta\xi$ and $\delta \mu$ are obtained by the first two diagonal 
elements of  $\Jm^{-1}(\thetav)$. It is easy to check that the CRB decreases as $O(1/\SNR)$ where $\SNR$ denotes the Signal-to-Noise Ratio at the receiver
\cite{kay-estimation}. 

\begin{figure}[ht]
\centerline{a)\includegraphics[width=8cm]{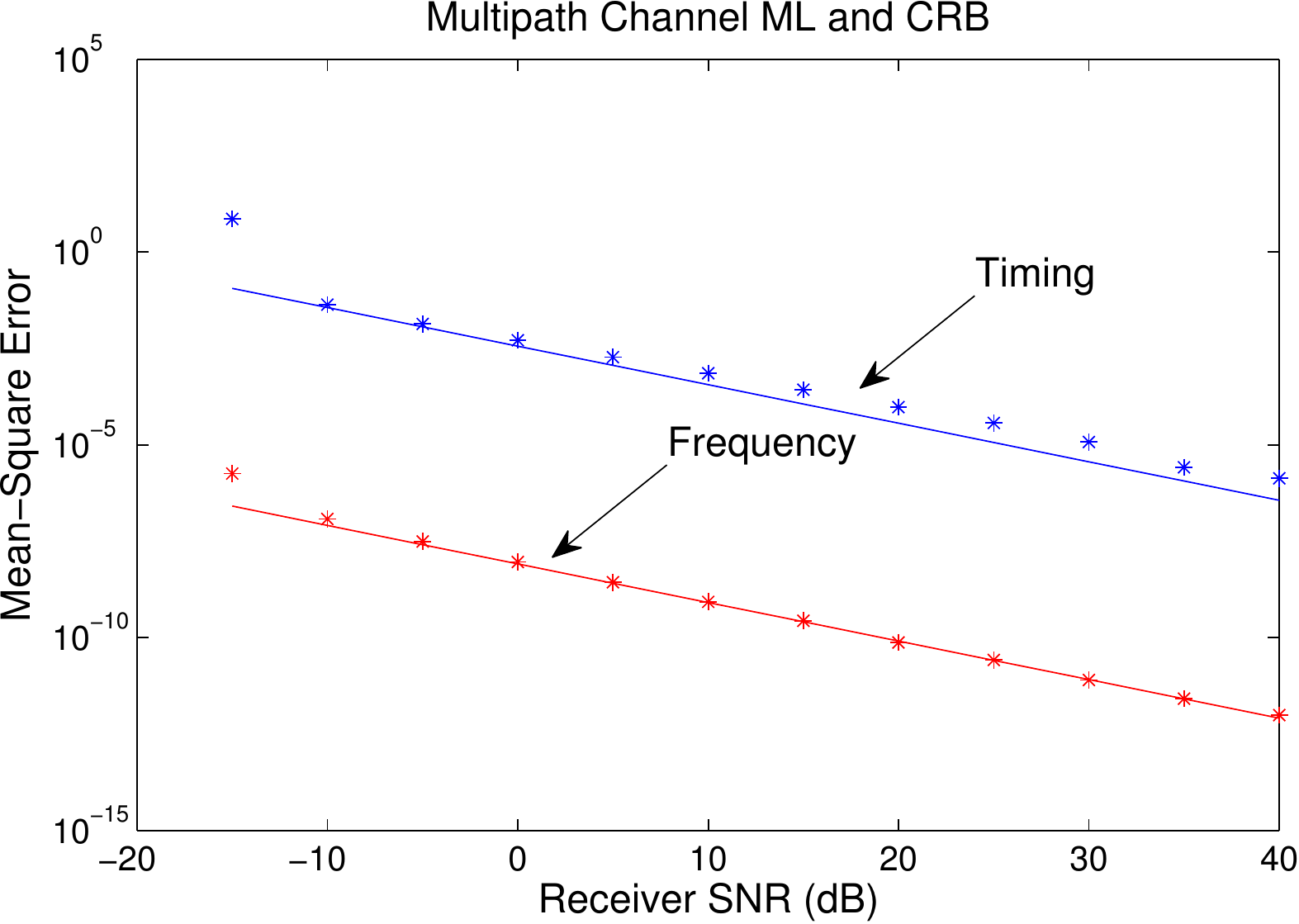} b)\includegraphics[width=8cm]{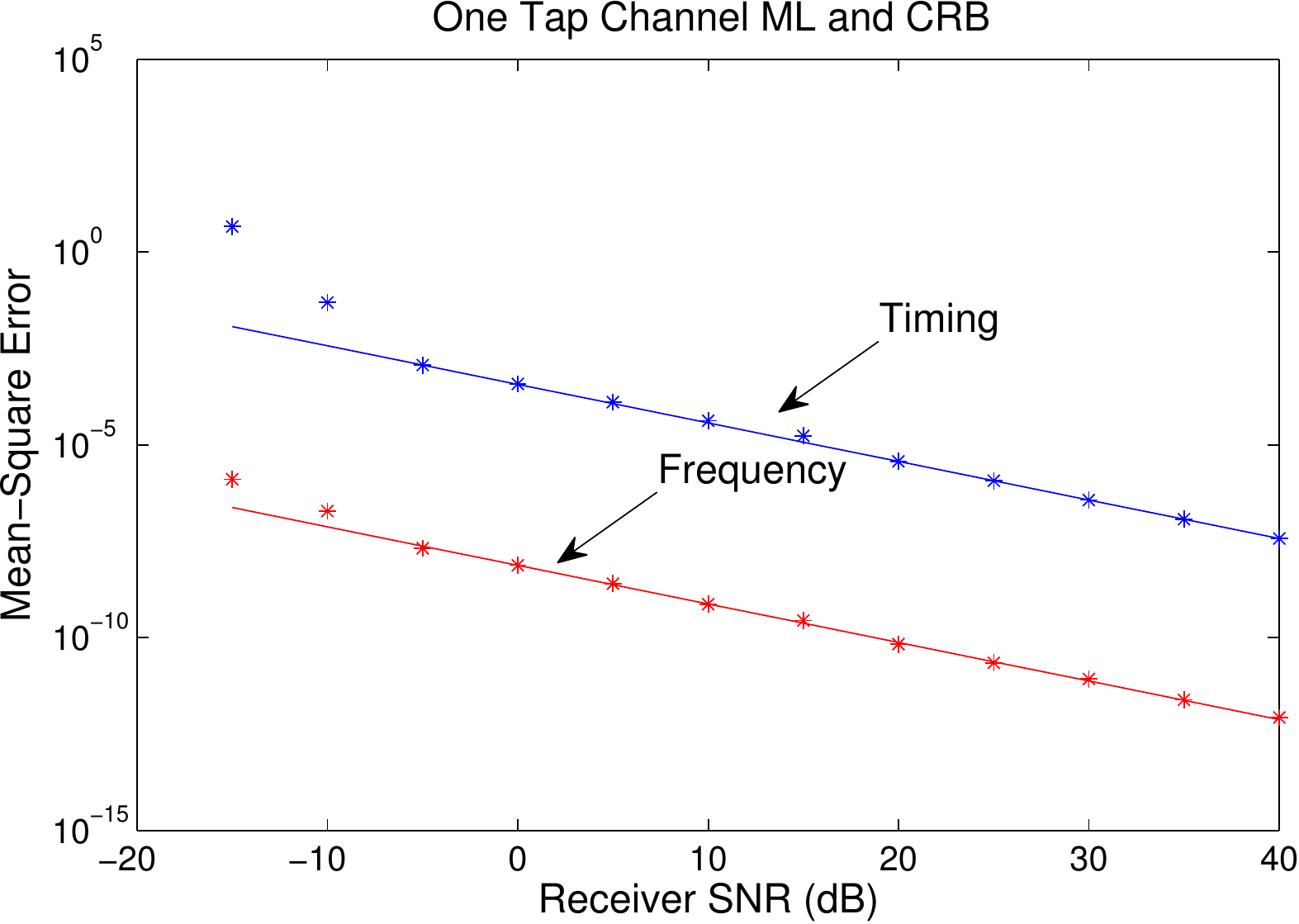}}
\caption{CRB (solid line) and simulated estimation MSE (asterisks) of the joint ML estimation a) for a multi-path channel example b) single-path channel example.}
\label{fig:mle}
\end{figure}

Fig.~\ref{fig:mle} a) shows the CRB for $\delta \xi = 0$ and $\delta \mu = 0$, 
$\hv = (1, 0.5, -0.2, 0.1, -0.05, -0.005)^\transp$, with path delays (normalized with respect to $T_s$) equal to
$(0, 1.75,    3.56,    7.90,   10.72,   15.30)$, respectively, and for a pilot sequence $\sv$ formed by 
a block of 64 pseudo-random chips with values in the QPSK constellation, repeated twice and separated by 
128 chips equal to 0, for a total pilot sequence length of 256 chips, corresponding to less than 4 64-carrier OFDM symbols (in fact, the OFDM symbols also include 
the cyclic prefix which here, for this time-domain pilot burst, are not necessary). 
The simulated MSE of the joint ML estimator derived before is also shown for comparison in the figure. 
As the figure reveals, the simulated estimation MSE is very close to the corresponding CRB even for very low SNR. 
Fig.~\ref{fig:mle} b)  shows the associated MSEs of the joint ML estimator assuming a single-path channel between the two APs. 
The behavior of the MSE is very similar to the case of Fig.~\ref{fig:mle} a), showing that the joint ML estimator is remarkably insensitive 
to the actual channel delay-intensity profile.

\bibliographystyle{IEEEtran}
\bibliography{IEEEabrv,refs}

\end{document}